%% file: bioarxiv.tex
\documentclass[a4paper]{article}
\pdfoutput=1
\usepackage[a4paper,top=3cm,bottom=2cm,left=3cm,right=3cm,marginparwidth=1.75cm]{geometry}
\usepackage[english]{babel}
\usepackage[affil-it]{authblk}
\usepackage[utf8x]{inputenc}
\usepackage[T1]{fontenc}

\usepackage[a4paper,top=3cm,bottom=2cm,left=3cm,right=3cm,marginparwidth=1.75cm]{geometry}
\usepackage{amsmath}
\usepackage{amssymb}
\usepackage{amsfonts}
\usepackage{fullpage}
\usepackage{bm}
\usepackage{bbm} 
\usepackage{bbold}
\usepackage{appendix}
\usepackage{graphicx}
\usepackage{xcolor}
\usepackage{tikz}
\usepackage{cite} %% Cite allows to have [1]
\usepackage{amsmath}
\usepackage{graphicx}
\usepackage{url}
\makeatletter
\renewcommand\@biblabel[1]{#1.} %from [1] to 1
\makeatother
\newcommand*{\myfont}{\fontfamily{phv}\selectfont}

\usepackage[colorinlistoftodos]{todonotes}
\usepackage[colorlinks=true, allcolors=blue]{hyperref}

\newcommand{\beginsupplement}{%
        \setcounter{table}{0}
        \renewcommand{\thetable}{S\arabic{table}}%
        \setcounter{figure}{0}
\renewcommand{\thefigure}{\Alph{figure}}%
     }

\usetikzlibrary{calc,decorations.markings}
\newcommand{\be}{\begin{equation}}
\newcommand{\ee}{\end{equation}}
\newcommand{\ba}{\begin{eqnarray}}
\newcommand{\ea}{\end{eqnarray}}
\newcommand{\lp}{\left(}
\newcommand{\rp}{\right)}

\usepackage{lineno}
\date{\scriptsize{\textsuperscript{*}To whom correspondence should be addressed. E-mail: barbarabravi@ymail.com, matthieu.wyart@epfl.ch}}
\title{Direct Coupling Analysis of Epistasis in Allosteric Materials}

\author[1,*]{Barbara Bravi}
\author[1]{Riccardo Ravasio} 
\author[2]{Carolina Brito}
\author[1,*]{Matthieu Wyart}

\affil[1]{Institute of Physics, \'Ecole Polytechnique F\'ed\'erale de Lausanne, CH-1015 Lausanne, Switzerland}
\affil[2]{Instituto de F\`isica, Universidade Federal do Rio Grande do Sul, CP 15051, 91501-970 Porto Alegre RS, Brazil}

\begin{document}
\maketitle

\begin{abstract} 
In allosteric proteins, the binding of a ligand modifies function at a distant active site. Such allosteric pathways can be used as target for drug design, generating  considerable interest in inferring them from sequence alignment data. Currently, different methods lead to conflicting results, in particular on the existence of long-range evolutionary couplings between distant amino-acids mediating allostery. Here we propose a resolution of this conundrum, by studying epistasis and its inference in models where an allosteric material is evolved \emph{in silico} to perform a mechanical task. We find in our model the four types of epistasis (Synergistic, Sign, Antagonistic, Saturation), which can be both short or long-range and have a simple mechanical interpretation. We perform a Direct Coupling Analysis (DCA) and find that DCA predicts well the cost of point mutations but is a rather poor generative model. Strikingly, it can predict short-range epistasis but fails to capture long-range epistasis, in consistence with empirical findings. We propose that such failure is generic when function requires subparts to work in concert. We illustrate this idea with a simple model, which suggests that other methods may be better suited to capture long-range effects. 
\end{abstract}

\vspace{1cm}
\noindent\fbox{%
    \parbox{0.9\textwidth}{%
    \textbf{Author summary}\\
     Allostery in proteins is the property of highly specific responses to ligand binding at a distant site. To inform protocols 
     of \emph{de novo} drug design, it is fundamental to understand the impact of mutations on allosteric regulation and whether it can 
     be predicted from evolutionary correlations. 
     In this work we consider allosteric architectures artificially evolved to optimize the cooperativity of binding at allosteric and 
     active site. We first characterize the emergent pattern of epistasis as well as the underlying mechanical phenomena, finding the four types of epistasis (Synergistic, Sign, Antagonistic, Saturation), which can be both short or long-range. The numerical evolution of these allosteric architectures allows us to benchmark Direct Coupling Analysis, a method which relies on co-evolution in sequence data to infer direct evolutionary couplings, in connection to allostery. We show that Direct Coupling Analysis predicts quantitatively point mutation costs but underestimates strong long-range epistasis. We provide an argument, based on a simplified model, illustrating the reasons for this discrepancy. Our analysis suggests neural networks as more promising tool to measure epistasis.
    }%
}

\section*{Introduction} 
Allosteric regulation in proteins allows for the control of functional activity by ligand binding at a distal allosteric site \cite{Guo2016} and its detection could guide drug design \cite{Dokholyan2016,Guarnera2016}. Yet, understanding  the principles responsible for allostery remains a challenge. How random mutations dysregulate allosteric communication  is a valuable information studied experimentally \cite{Tang2017} and computationally \cite{Ahuja2017}. Several analyses have highlighted the non-additivity of mutational effects or \emph{epistasis}. This ``interaction'' between mutations can span long-range positional combinations \cite{olson2014}, results in either beneficial or detrimental effects to fitness \cite{devisser2011}, and shapes protein evolutionary paths \cite{Starr2016}. Given the combinatorial complexity of its characterization, empirical patterns of epistasis are still rather elusive \cite{Ortlund2007,Natarajan2013,Krug2013,Salinas18}.
Concomitantly, progress in sequencing has led to an unprecedented increase of availability of data arranged into Multiple Sequence Alignments (MSAs) \cite{Durbin1998} containing many realizations of the same protein in related species. Different methods have been developed to extract information from sequence variability, e.g.\ Statistical Coupling Analysis \cite{Ranghanatan2003, Ranghanatan2011} was applied to allostery detection in proteins. It was argued that the allosteric pathway was encoded in spatially extended and connected {\it sectors}, groups of strongly co-evolving amino-acids, supporting that long-range information on the allosteric pathway is contained in the MSA. Another approach, Direct Couplings Analysis (DCA) \cite{Morcos11}, aims at inferring evolutionary couplihngs between amino-acids. Direct couplings predict successfully residue contacts \cite{Morcos11} so to inform the discovery of new folds \cite{Baker2017}, allow one to describe evolutionary fitness landscapes \cite{ferguson2013, mann2014, Charlaix2016, Figliuzzi2016, hopf2017} and correlate with epistasis \cite{nelson18, Poelwijk2018}. In the context of allostery, there is no statistical evidence for the existence of long-range direct couplings that would reveal allosteric channels \cite{Anishchenko2017}, in apparent contradiction with the existence of extended sectors reported in \cite{Ranghanatan2011} and the observation of long-range epistasis \cite{olson2014}. It is therefore an open question why a pairwise model should be successful at predicting protein structure, but not long-range functional dependencies.
In this work we propose an explanation for this discrepancy, by benchmarking DCA in models of protein allostery where a material evolves \emph{in silico} to achieve an ``allosteric'' task \cite{Hemery15,Rocks17, Flechsig17, Yan17, Yan18,Tlusty16,Dutta2017}. We consider recent models incorporating elasticity \cite{Rocks17, Flechsig17, Yan17, Yan18,Dutta2017}, in which long-range co-evolution \cite{Yan17}, elongated sectors \cite{Yan17} and long-range epistasis \cite{Dutta2017} are present and can be interpreted in terms of the propagation of an elastic signal \cite{Dutta2017}. We focus on materials evolved to optimize cooperative binding over large distances \cite{Yan18}, and find that the four types of epistasis (Synergistic, Sign, Antagonistic, Saturation) exist over a wide spatial range. We perform DCA and find that it predicts well the cost of point mutations but is a rather poor generative model. It can predict short-range epistasis but fails to capture long-range effects, in agreement with empirical findings \cite{Anishchenko2017}. Moreover, we test this result for one allosteric protein, the PDZ domain, where epistasis was experimentally measured in \cite{Salinas18} along with the inference of DCA energetic couplings, showing support for our prediction. We illustrate why it may be so via a simple model, which suggests that neural networks may be better suited than DCA to capture long-range effects.
 
\subsection*{Model for the evolution of allostery}
\begin{figure}%[tbhp]
\centering
\includegraphics[width=.7\linewidth]{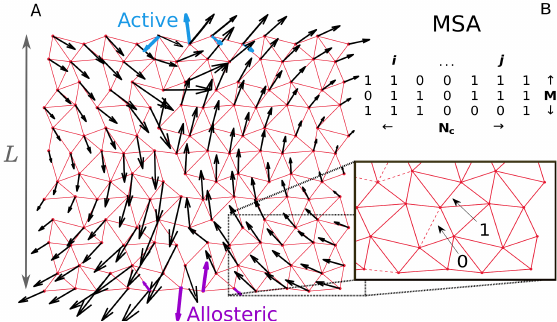}
\caption{\textbf{Study of co-evolution in artificial allosteric networks}. A: Example of an elastic network made of harmonic springs (red) evolved \emph{in silico} to maximize the cooperativity between the allosteric site (purple) and the active site (blue). The response to binding at the allosteric site is indicated by black arrows, and is found to follow a shear motion. B: Each network corresponds to a sequence of 0 and 1 coding for the spring absence or presence. Our scheme allows us to generate a large number $M$ of such sequences, each corresponding to a slightly different shear architecture.}
\label{fig:msa}
\end{figure}

We follow the scheme of \cite{Yan17,Yan18} where a protein is described by an elastic network of size $L$ made of harmonic springs of unit stiffness (here we consider $L=12$). Binding events are modeled as imposed displacements either at the ``allosteric'' or at the ``active'' site (each consisting of several nodes), as shown in color in Fig.\ \ref{fig:msa}A. Such imposed displacements elicit an elastic response in the entire protein and cost some elastic energy, which defines our binding energy (see Sec.\ \ref{sec:mech_epi_app} in S1 Text). Following \cite{Yan18}, the fitness ${\cal F}$ measures the cooperativity of binding between allosteric and active site and is defined as the energy difference ${\cal F} \equiv  E^{{\cal A}c} - (E^{{\cal A}c,{\cal A}l} - E^{{\cal A}l})$ where $E^{{\cal A}c}$, $E^{{\cal A}l}$ and $E^{{\cal A}c,{\cal A}l}$ are respectively the elastic energy of binding at the active site only (${\cal A}c$), at the allosteric site only (${\cal A}l$) and at both sites simultaneously (${\cal A}c,{\cal A}l$). In the limit of weak elastic coupling between allosteric and active site, the fitness can be rewritten approximately as (see Sec.\ \ref{sec:mech_epi_app} in S1 Text)

\be
\label{eq:fit_res}
{\cal F} \approx {\bf F}^{{\cal A}c}\cdot\bm{R}^{{\cal A}l\rightarrow {\cal A}c}
\ee
where ${\bf F}^{{\cal A}c}$ is the force field imparted by substrate binding on the nodes of the active site, and $\bm{R}^{{\cal A}l \rightarrow {\cal A}c}$ is the displacement field induced at the active site by ligand binding. The product ${\bf F}^{{\cal A}c}\cdot\bm{R}^{{\cal A}l\rightarrow {\cal A}c}$ is an estimate of the change of mechanical work required for binding the substrate at the active site caused by binding the ligand at the allosteric site. Note that each field in Eq.\ \ref{eq:fit_res} is of dimension $n_0 d$, where $n_0=4$ is the number of nodes in the active site and $d=2$ the spatial dimension. 

Such networks are evolved by changing the position of springs according to a Metropolis-Monte Carlo routine to maximize ${\cal F}$. At each step, the fitness difference with respect to the previous configuration $\Delta {\cal F}$ is computed and the new configuration is accepted with a probability $p = \min(1, \exp{\beta \Delta{\cal F}})$. $\beta$ is an evolution inverse temperature controlling the selection pressure for high fitness ${\cal F}$, we choose $\beta=10^4$ as at this temperature networks probed have the highest fitness our protocol can reach \cite{Yan18}. We sample every 1000 time steps after an initial equilibration time of $10^5$ steps. At long times one obtains 
a cooperative system of typical ${\cal F} \sim 0.2$, whose architecture depends on the spatial dimension and boundary conditions \cite{Yan18}. Here we consider a network in $d=2$ dimensions with periodic boundaries, equivalent to a cylindrical geometry, where the response to binding evolves towards a \emph{shear} mode (see Fig.\ \ref{fig:msa}A). With our scheme we can generate thousands of networks with a similar design. A sequence $\bm{\sigma}$ of 0 and 1, where $\sigma_i =1$ stands for the presence of a spring at link $i$ and $\sigma_i =0$ for its absence, can be associated to any network, leading to a Multiple Sequence Alignment (MSA) of networks performing the same function (see Fig.\ \ref{fig:msa}B).

\section*{Results}

\subsection*{Nature and classification of epistasis}
\label{sec:char_epi}
The cost of a single mutation (i.e.\ changing the occupancy) at some link $i$ is defined as $\Delta {\cal F}_i = {\cal F} - {\cal F}_i$ where ${\cal F}$ is the original fitness and ${\cal F}_i$ the one of the network after the mutation. Single mutation costs $\Delta {\cal F}_i$ are expected to be positive since the original network has been selected to have close-to-maximal fitness.

We denote by \ $\Delta {\cal F}_{ij} = {\cal F} - {\cal F}_{ij}$ the cost of a double mutation at $i$ and $j$. Epistasis between loci $i$ and $j$ is then defined as $\Delta \Delta {\cal F}_{ij}\equiv \Delta {\cal F}_{ij} - \Delta \mathcal{F}_{i} - \Delta \mathcal{F}_{j}$. We find that generically,  the dominant effect of mutations  is to affect the propagation of the signal $\bm{R}^{{\cal A}l \rightarrow {\cal A}c}$, which depends on 
the arrangement of links in the network. In general, mutations do not affect how binding at the active site locally generates force, as shown in Sec.\ 1 in S1 Text. Using this observation and following Eq.\ \ref{eq:fit_res}, epistasis follows approximately
\begin{align}
\label{eq:epi_coop_main}
\Delta \Delta{\cal F}_{ij} \approx -  {\bf F}^{{\cal A}c} \cdot \Big(\delta \bm{R}_{ij}^{{\cal A}l \rightarrow {\cal A}c} - \delta \bm{R}_{i}^{{\cal A}l \rightarrow {\cal A}c} - \delta \bm{R}_{j}^{{\cal A}l \rightarrow {\cal A}c} \Big) \notag
\end{align}
where\ $\delta {\bf R}^{{\cal A}l \rightarrow {\cal A}c}_{i} ={\bf R}^{{\cal A}l \rightarrow{\cal A}c}_{i} - {\bf R}^{{\cal A}l \rightarrow{\cal A}c}$, and ${\bf R}^{{\cal A}l \rightarrow {\cal A}c}_{i}$ is the allosteric response at the active site of the protein mutated at  link $i$.  $\delta {\bf R}^{{\cal A}l \rightarrow {\cal A}c}_{j}$ and $\delta {\bf R}^{{\cal A}l \rightarrow {\cal A}c}_{ij}$ follow analogous definitions. We denote by  $\theta$  the angle between $\delta {\bf R}^{{\cal A}l \rightarrow {\cal A}c}_{i}$ and $\delta {\bf R}^{{\cal A}l \rightarrow {\cal A}c}_{j}$.

\begin{figure}
\centering
\includegraphics[width=0.7\linewidth]{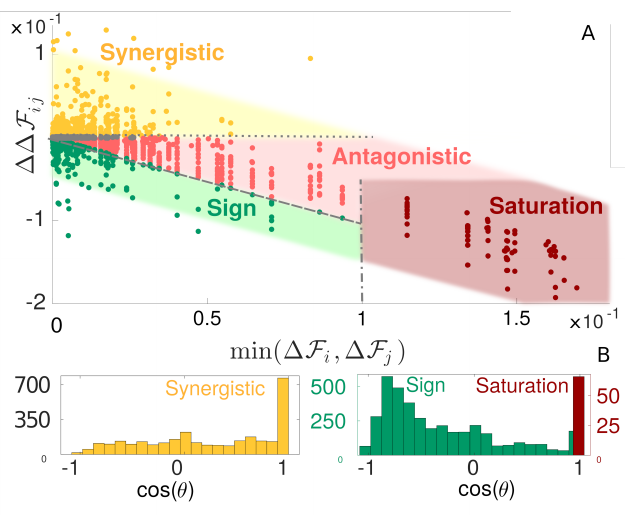}
\caption{\textbf{Classification and mechanical characterization of epistasis in our model of allosteric cooperativity.}
A: Phase diagram of epistasis in our allosteric material. All quantities are averages over $50$ configurations obtained in a single run. 
The shaded area is taken with arbitrary width and a -1 slope as a guide to the eye. We show the lines $\Delta \Delta {\cal F}_{ij}  = 0$ (dotted style), which corresponds to no epistasis (and divides synergistic from antagonistic/sign epistasis), $\Delta \Delta {\cal F}_{ij} = \max(\Delta \mathcal{F}_i, \Delta \mathcal{F}_j)$ (dashed style), separating sign and antagonistic epistasis, and $\min(\Delta \mathcal{F}_i, \Delta \mathcal{F}_j) = 0.1$ (dash-dotted style), the threshold set to distinguish lethal mutations (corresponding to the saturation region). Points in grey correspond to epistasis $ < 5 \times 10^{-4}$ and are excluded from our analysis. B: Histograms of $cos(\theta)$ for synergistic, sign and saturation epistasis.}
\label{fig:epistasis}
\end{figure}

Consider the case where the cost of a double mutation is dominated by the strongest point mutation, i.e.\ $\Delta {\cal F}_{ij}\approx \max(\Delta {\cal F}_i,\Delta {\cal F}_j)$. It leads to:
\be
\label{eq:scalingddf}
\Delta \Delta {\cal F}_{ij}  \approx -\min(\Delta \mathcal{F}_i, \Delta \mathcal{F}_j).
\ee
%This assumption does capture a significant part of epistasis, 
Interestingly, this situation does capture the main trend of epistasis in our data, especially when it is strong, as shown in Fig.\ \ref{fig:epistasis}A (see dashed line). This observation suggests to classify pairs of loci in terms of their epistasis and the minimal associated mutation cost $\min(\Delta \mathcal{F}_i, \Delta \mathcal{F}_j)$ as performed in Fig.\ \ref{fig:epistasis}A. First of all, no epistasis corresponds to purely additive mutations, i.e.\ 
$\Delta \Delta {\cal F}_{ij}=0$, see dotted line in Fig.\ \ref{fig:epistasis}A. Next, we observe the following regimes

{\it Saturation:} We define  mutations with $\Delta {\cal F} > 0.1$ as "lethal". This somewhat arbitrary definition corresponds to  $50\%$ of loss of fitness. Pairs of such lethal mutations (which represent $\sim 0.1\%$ of all pairs, a sparsity in line with experimental findings \cite{Poelwijk2018}) have the strongest epistasis in absolute value, and follow closely Eq.\ \ref{eq:scalingddf}, as visible in Fig.\ \ref{fig:epistasis}A. Physically, these mutations essentially shut down signal propagation by themselves with ${\bf R}^{{\cal A}l \rightarrow{\cal A}c}_{i}\approx {\bf R}^{{\cal A}l \rightarrow{\cal A}c}_{j}\approx 0$, in such a way that the double mutation has the effect of a single one with ${\bf R}^{{\cal A}l \rightarrow{\cal A}c}_{ij}\approx 0$. This view is confirmed 
in Fig.\ \ref{fig:epistasis}B by the observation that $\cos(\theta)\approx 1$, as follows from $\delta {\bf R}^{{\cal A}l \rightarrow {\cal A}c}_{i} \approx \delta {\bf R}^{{\cal A}l \rightarrow {\cal A}c}_{j}\approx  - {\bf R}^{{\cal A}l \rightarrow{\cal A}c}$. Saturation is then a form of very high ``diminishing-returns'' epistasis, for which evidence from data and support from theoretical models are accumulating \cite{Chou2011,Krug2016}.

\emph{Antagonistic.} Further up along the diagonal of Eq.\ \ref{eq:scalingddf} in Fig.\ \ref{fig:epistasis}A, this saturation effect becomes milder. It is more akin to ``antagonistic'' epistasis \cite{Desai2007, devisser2011}, whereby, after a first mutation, making a second one results only in a weak additional change. Antagonistic epistasis is also known as positive magnitude epistasis (where positivity indicates that the double mutant is fitter than expected from the additive case).

\emph{Sign.} In the intermediate range of mutation costs with $\min(\Delta \mathcal{F}_i, \Delta \mathcal{F}_j)<0.1$, more compensatory epistatic interactions can take place, where the fitness cost of a deleterious mutation is diminished by the second mutation (i.e.\
$\Delta {\cal F}_{ij} < \max(\Delta \mathcal{F}_i, \Delta \mathcal{F}_j)$). Thus some mutations can become beneficial (i.e.\ increase the fitness) in presence of another mutation, 
and this resembles the ``sign'' epistasis empirically detected \cite{devisser2011, Lalic2012}. Geometrically, it corresponds to situations where the two mutations deform the signal in opposite directions, so the second one can partially re-establish fitness. In support of this, Fig.\ \ref{fig:epistasis}B shows that for sign epistasis $\cos(\theta)$ tends to be negative. 

\emph{Synergistic.} Positive-sign values of $\Delta \Delta {\cal F}_{ij}$ indicate ``synergistic'' epistasis. It occurs if two mutations perturb the elastic signal in the same direction, causing more damage than expected if they were purely additive. As clear from Fig.\ \ref{fig:epistasis}B, $\cos(\theta)$ tends to be positive in this case.

\subsection*{Direct Coupling Analysis} 
We evolve numerically $M$ configurations maximizing cooperativity ${\cal F}$, each yielding a realization of a (variable) shear design. We sample a configuration for every initial condition to avoid introducing a bias in the sampling due to their high similarity. (We thus eliminate the possibility of our sequences to display ``phylogenetic'' effects, i.e.\ correlations due to a common evolutionary history, known to complicate the inference from sequence data and to require \emph{ad hoc} corrections, see e.g.\ \cite{Review_protein_seq}). We find that the average Hamming distance among the obtained sequences is $\sim 20\%$ of their length. Our set of sequences is analogous to a protein MSA – importantly, in this analogy the role of an amino-acid is played by a link, which can be stiff ($\sigma_i=1$) or not ($\sigma_i=0$, no springs). In practice we take $M = 135 000$, much larger than the sequence length $N_c= (3L^2 - 2L)=408$. Working in such an over-sampling regime (which is generally not the case for real proteins) ensures that the  limitations of the inference we find below are not due to sampling, but to the model underlying DCA.

\begin{figure}[t]
\centering
\includegraphics[width=.7\columnwidth]{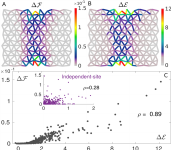}
\caption{\textbf{Prediction of mutation costs by DCA}. Maps of true $\Delta{\cal F}$ (A) and DCA-inferred $\Delta {\cal E}$ (B) single mutation costs, averaged over $1.5\times10^3$ configurations randomly chosen from the MSA. Their patterns are very similar, revealing high costs near the allosteric and active sites and in the shear path connecting them. C: Scatter plot showing the strong correlation between $\Delta{\cal F}$ and $\Delta {\cal E}$ for all links (averaged over $1.5\times10^3$ configurations). The estimation of mutation costs based on an independent-site model (i.e.\ on conservation) correlates poorly with the true cost (inset), proving the need for incorporating correlations for proper prediction of mutation costs. The correlation is quantified via the Pearson correlation coefficient, $\rho$.}
\label{fig:mutations_coop1}
\end{figure}

Next, for a statistical analysis of these sequences, we use DCA, which is based on the idea of fitting the observed single-site $\langle \sigma_i \rangle = 1/M\sum_m \sigma_{i}^m $ and pairwise $\langle \sigma_i \sigma_j\rangle = 1/M\sum_m \sigma_{i}^m \sigma_{j}^m$ frequencies of links by the probability distribution $ P(\bm{\sigma})$ with maximal entropy (as this ensures the least biased fit of data under such empirical constraints). In our setup this approach leads to

\ba
\label{eq:BG}
 P(\bm{\sigma})&=&\frac{1}{Z}\exp{(-{\cal E}(\bm{\sigma}))}\\
{\cal E}(\bm{\sigma}) &=& -\sum_{i< j}J_{ij}\sigma_i \sigma_j - \sum_i h_i \sigma_i
\label{eq:energy}
\ea
which is equivalent to an Ising model where $\sigma_i = 0,1$ would denote the two states (down, up) of spins. In this setting, ${\cal E}$ is an estimation of $\beta {\cal F}$, $\beta$ being the inverse evolution temperature. In all the comparisons (e.g.\ Fig.\ \ref{fig:mutations_coop1}) we omit $\beta$ as we are only interested in testing the proportionality between ${\cal E}$ and ${\cal F}$. The ``fields'' $h_i$ and ``couplings'' $J_{ij}$ are inferred to match $\langle \sigma_i \rangle$ and $\langle \sigma_i \sigma_j \rangle$. The inference of these parameters can be performed with several algorithms, we focus on ACE (Adaptive Cluster Expansion) \cite{Cocco11, Cocco12}, an approximate technique developed from statistical physics ideas, combined with maximum likelihood, an exact technique. This approach is extremely accurate and we compare it to a method more approximate, but much faster computationally, as mean field Direct Coupling Analysis (mfDCA) \cite{Morcos11}, 
see Methods for details on the implementation.

In this way we can benchmark DCA in the context of allosteric materials and test if it: (i) reproduces accurately the cost of single mutations; (ii) is a good generative model, i.e.\ if it can generate new sequences with high fitness and (iii) can predict epistasis.

\subsubsection*{Inferring mutation costs} 
Fig.\ \ref{fig:mutations_coop1}A shows the map of true mutation costs, indicating a large cost near the allosteric and active sites as well as in the central region where the allosteric response displays high shear (as documented in \cite{Yan18}). DCA enables one to infer this map by computing the estimated mutation cost $\Delta {\cal E}_i = {\cal E}_i - {\cal E}$ for a mutation at a generic link $i$, Fig.\ \ref{fig:mutations_coop1}B. 
The comparison is excellent, as evident also from the high correlation revealed by the scatter plot Fig.\ \ref{fig:mutations_coop1}C. Importantly, including pairwise couplings is key for inferring mutation costs, as a model based on conservation alone (a standard measure of mutation costs, see Methods) performs poorly in this case, see inset of Fig.\ \ref{fig:mutations_coop1}C. 

\subsubsection*{Generative power of DCA}
\begin{figure}[t]
\centering
\includegraphics[width=0.7\columnwidth]{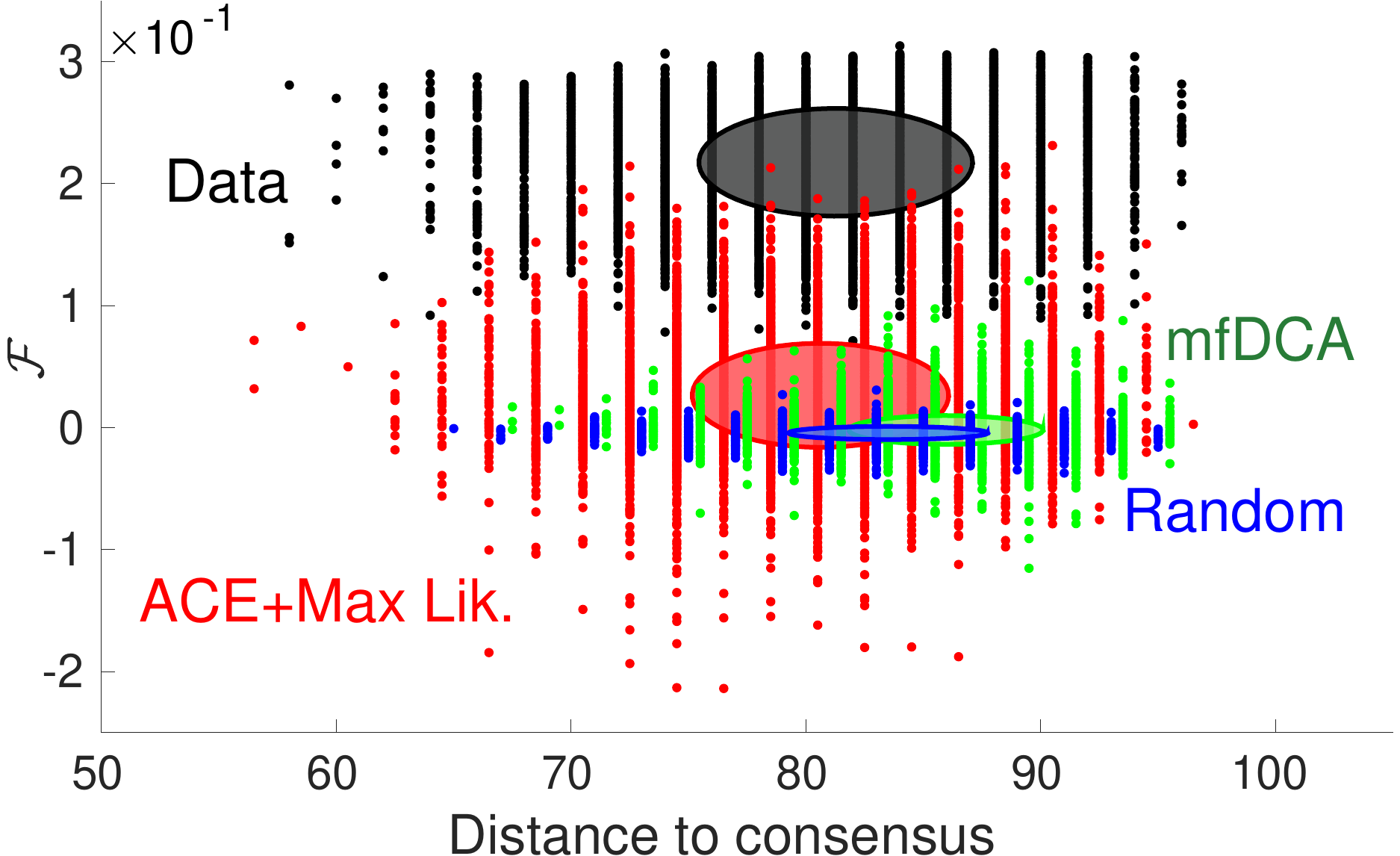}
\caption{\textbf{Generative performance of DCA}. Fitness vs distance to consensus of configurations generated by the inferred model, following the representation of \cite{Jacquin2016}. The sampling is done from $P(\bm{\sigma})$ of Eq.\ \ref{eq:energy} (a Boltzmann-Gibbs probability distribution), whose parameters have been inferred via ACE + maximum likelihood (red cloud) or mfDCA (green cloud). Original high fitness configurations (black cloud) and random ones (blue) are added as a reference. Each cloud consists of $10^4$ sequences and the drawn ellipse gives one standard deviation around the mean in both horizontal and vertical directions. Distances to consensus of ACE + maximum likelihood, mfDCA and random sequences are shifted by respectively $+0.7$, $-0.7$ and $-1.3$ for better visibility.
}
\label{fig:generative}
\end{figure}

Once the model of 
Eqs.\ \ref{eq:BG}, \ref{eq:energy} is inferred, can it be used to generate new sequences with a high fitness, as previously shown for models of protein folding \cite{Jacquin2016}? To answer this question, we generate new sequences by Monte Carlo sampling from the probability distribution Eq.\ \ref{eq:BG}. Fig.\ \ref{fig:generative} shows the fitness of the obtained sequences vs their distance to ``consensus'' - the consensus being the most representative sequence of the MSA, i.e.\ where springs occupy the positions with largest mean occupancy. We find that (i) the variability of the MSA, quantified by the distance to consensus, is well reproduced (ii) the fitness is much more variable than for random sequences, with a few sequences that do perform as well as evolved ones (which never occurs for random sequences) but (iii) the mean obtained fitness is rather low, although  larger, in a statistically significant way, than the one of random configurations (which is zero). As shown in Fig.\ \ref{fig:generative}, these results deteriorate further if a more approximate algorithm as mfDCA is used to infer parameters. We have checked that the generative performance is not improved by lowering the temperature of the Monte Carlo sampling. Overall, these results suggest that the generative power of DCA is limited in the context of allostery, in contrast with results for models of protein folding \cite{Jacquin2016}. Thus an Ising model, a quadratic model accounting for conservation and correlations in the MSA (first and second order statistics), although it can capture some features of the shear design (e.g.\ the inhomogeneous distribution of coordination, as shown in Fig.\ \ref{fig:prop_gen_coop}), is a rather drastic approximation for the actual allosteric fitness. Indeed we have tested that higher orders as the third moment are not well reproduced (see Fig.\ \ref{fig:DCA}), suggesting that the longer-range correlations induced by allostery are not well captured by a pairwise model. On the other hand, for protein structure predictions, several works as \cite{Figliuzzi2018} suggest that local correlations between residues in spatial contact are well-captured by a pairwise model, even beyond pairwise correlations. 
To test our findings, it would be interesting to condition the analysis of e.g.\ \cite{Figliuzzi2018} on the distance between residues considered and see if the 3-body correlations are still captured when the residues are further apart. It would also be relevant to restrict the study to allosteric proteins only, to check whether statistical properties are changed, in such a way as to gauge the effect of allosteric vs folding constraints in proteins.

In what follows we shall emphasize in particular the failure of DCA to infer long-range epistasis.

\begin{figure}
\centering
 \includegraphics[width=0.7\linewidth]{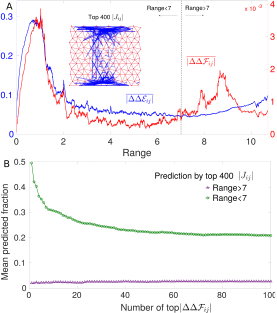}
\caption{\textbf{Prediction of epistasis by DCA}. A: Running average of the absolute value of epistasis $\Delta\Delta{\mathcal F}_{ij}$ and of DCA prediction $\Delta\Delta{\mathcal E}_{ij}$ for $1.5\times10^3$ configurations as a function of the distance between link $i$ and $j$. The trends are nearly identical at short distances but at long distance DCA underestimates epistasis. Inset: Top 400 inferred couplings. They are mostly short range with only a few long-range couplings connecting the allosteric and the active site. Next we assess the prediction of epistasis in single configurations by these top 400 couplings. We consider separately long-range (> 7) and short-range (< 7) pairs of links, and rank them respectively in terms of the epistasis magnitude $|\Delta\Delta{\mathcal F}_{ij}|$. B shows which fraction of these pairs - averaged over $100$ configurations randomly chosen - belongs to the 400 largest couplings, as a function of the number of pairs with maximal epistasis considered. Clearly coupling magnitude has less predictive power at large distances than at short ones. The random expectations for these mean predicted fractions are $0.0041$ for short-range pairs and $0.0009$ for long-range ones (they are both significantly lower than the values reported here).
This feature stays robust also if we increase, e.g.\ up to 1000, the number of top couplings for prediction (see Fig.\ \ref{fig:ppv_min_coop}A).} 
\label{fig:ddf_vs_J}
\end{figure}

\begin{figure}
\begin{center}
\includegraphics[width=0.8\textwidth]{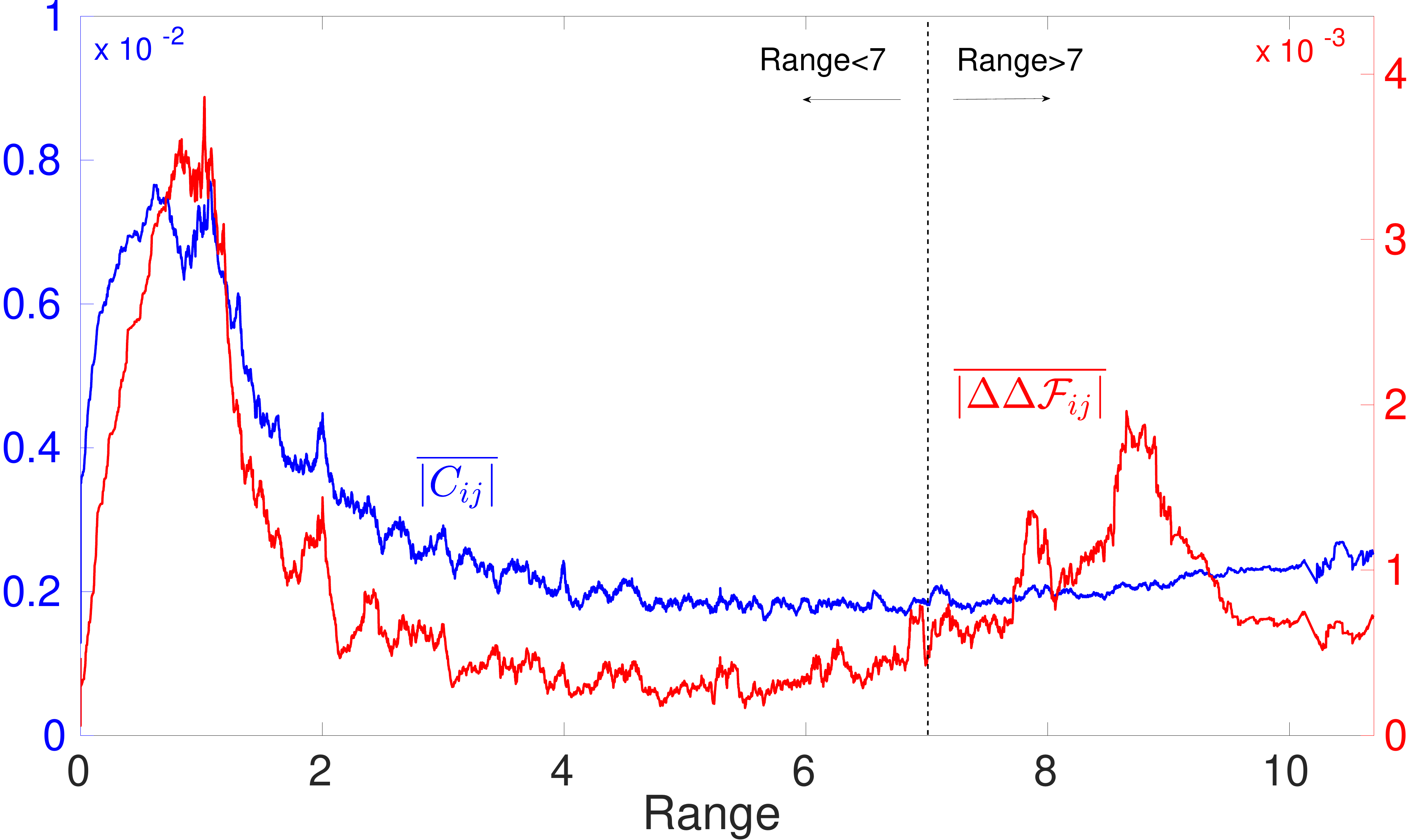}
\end{center}
\caption{Running average of the absolute value of connected correlations $C_{ij} = \langle \sigma_i \sigma_j \rangle - \langle \sigma_i \rangle \langle \sigma_j \rangle$ and of epistasis $\Delta\Delta{\mathcal F}_{ij}$ for the same $1.5\times10^3$ configurations of Fig.\ \ref{fig:ddf_vs_J}A as a function of the distance between link $i$ and $j$.}
\label{fig:epi_corr}
\end{figure}

\subsection*{Inferring epistasis with DCA}  
From Eq.\ \ref{eq:energy} one readily has that the DCA prediction for epistasis follows $\Delta \Delta {\cal E}_{ij} = -J_{ij}(2\sigma_i-1)(2\sigma_j-1)$, implying $|\Delta \Delta {\cal E}_{ij}|=|J_{ij}|$. Hence, within DCA, the epistasis magnitude is simply the one of evolutionary couplings. In the inset of Fig.\ \ref{fig:ddf_vs_J}A  we show the spatial location of the top 400 pairs of links with highest coupling magnitude, illustrating that long-range couplings are rare. Yet, as implied jointly by Fig.\ \ref{fig:epistasis}A (showing that pairs of sites with large mutation cost systematically display strong epistasis) and Fig.\ \ref{fig:mutations_coop1}A (showing that sites with a large mutation cost can be distant), long range epistasis is present in our model, meaning that DCA fails to capture it. This fact is demonstrated quantitatively in Fig.\ \ref{fig:ddf_vs_J}A showing the mean epistasis $|\Delta \Delta {\cal F}_{ij}|$ and mean DCA prediction $|\Delta \Delta {\cal E}_{ij}|$ as a function of distances. The DCA-predicted trend 
reproduces the original one at small distances but strongly underestimates long-range epistasis. This is further evidenced in Fig.\ \ref{fig:ddf_vs_J}B showing that the average fraction of long-range pairs (range > 7) with the largest epistasis which falls in the list of the 400 pairs with largest couplings is much smaller than for short-distance pairs (< 7). However, even at short distance the prediction by $|J_{ij}|$ is not excellent but it is remarkably improved if, as done in \cite{Salinas18,Poelwijk2018}, one considers epistasis averaged over several configurations (see Sec.\ \ref{sec:epi_SI} in S1 Text). (This result is in contrast to the remarkable performance of DCA in residue contact prediction, which guided the discovery of novel protein structures \cite{Baker2017}. We recall that couplings inferred by the most accurate DCA algorithms exhibit maximal precision (i.e. number of true predicted contacts divided by the total number of predictions equal to 1) up to a number of contacts comparable with the protein size \cite{Barton16,Ekeberg2013}). Our finding is consistent with the lack of empirical evidence for long-range inferred couplings in allosteric proteins \cite{Anishchenko2017}.

To better investigate the reasons for this phenomenon  in our \emph{in silico} model, we report evolutionary correlations as a function of distance in Fig.\ \ref{fig:epi_corr}. We find that, although strong long range epistasis occurs, large long-range correlations are absent (a fact in some sense more surprising that not finding long-range couplings, since in principle short-range couplings alone could result in long-range correlations). The absence of long-range correlations suggests that it will be particularly challenging to capture long-range functional dependencies from low order statistics of the MSA alone. Consistently with this observation, statistical approaches based on principal components of the MSA covariance such as Sectors \cite{Lockless99, Halabi09} or Inverse Covariance Off-Diagonal (ICOD) \cite{Bitbol19} do not lead overall to better predictions of epistasis in our context, as we show in S1 Text, Sec.\ \ref{sec:epi_sca}. Among these approaches, we find that the best predictor of long-range epistasis is ICOD, a result that would be interesting to benchmark also in other systems. 

\subsubsection*{A proposed explanation for the failure of DCA at long-distances} We propose that the failure of DCA at long-range stems from its inability to describe a function that requires many subparts of the system to work in concert, when each subpart can be of different type. For example, in allosteric proteins on short length scales  soft regions must exist where shear propagates \cite{Yan18,Mitchell16}, giving rise to local constraints. Yet, the exact location of these soft regions can vary in space. On a larger length scale, these regions must assemble to create an extended soft elastic mode  \cite{Rios05,Zheng06,Yan18}, which generates global constraints: for the shear architectures it implies the presence of a soft path between the allosteric and active site, whose position however can fluctuate. 

\begin{figure}
\centering
\includegraphics[width=0.7\linewidth]{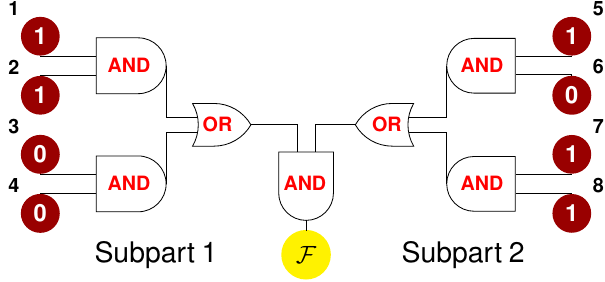}
\caption{\textbf{Sketch of a simple model for protein function}. A system is arranged into 2 subparts which must work jointly to accomplish a given function (AND gate). Each subpart is composed of 2 groups, i.e.\ can be of 2 types (OR gate), to work each type must satisfy some constraints (AND gate between single units).}
\label{fig:logic_gate}
\end{figure}

We argue that when applied to systems whose function is organized in such a hierarchical way, DCA underestimates long-range constraints. To illustrate this point, we introduce a Boolean model, shown in Fig.\ \ref{fig:logic_gate}. A generic ``function'' is achieved by two subparts that must work in concert (AND gate) and that can be of two different types (OR gate) but each must be functional (AND gate). This model comprises 8 units, taking the value 0 or 1, decomposed into 4 groups: 2 groups are the possible types of subpart 1 (left in Fig.\ \ref{fig:logic_gate}) and the other 2 the possible types of subpart 2 (right). A configuration is ``functional'' if 2 units of the same group are simultaneously in state 1 for each subpart. There are  49 functional configurations, whose fitness is fixed to ${\cal F}$, all other configurations have fitness $0$. We assume that ${\cal F}$ is large in such a way that the sequences in the MSA are only the 49 functional ones, with a uniform distribution. It is straightforward to calculate epistasis in this model, as well as single-site and pairwise frequencies from which couplings $J_{ij}$ and fields $h_i$ can be inferred. In particular we can compare $\Delta\Delta {\cal F}_{ij}$ and $\Delta\Delta {\cal E}_{ij}$ for units $i$ and $j$ either in the same group (or in the same subpart), so locally constrained by function (at ``short distance'', e.g.\ $i=1$ and $j=2$), or in the two different subparts, thus globally constrained (at ``long distance'' e.g.\ $i=1$ and $j=5$). 
We obtain (see Sec.\ \ref{sec:toy_SI} in S1 Text) that $|\Delta\Delta {\cal F}_{12}|/|\Delta\Delta {\cal F}_{15}|\approx 2.3$: global and local constraints lead to relatively similar short range and long-range epistasis. Yet we find that epistasis between subparts is noticeably underestimated by DCA in contrast to epistasis within subparts. To show this, we look at the DCA prediction for the ratio of epistasis between two pairs of sites divided by the true ratio of epistasis. For pairs of sites belonging to the same subpart, DCA predicts equally well epistasis. For example, considering the pair of sites (1,2) and the pair (1,3), one finds  $|\Delta\Delta {\cal E}_{13}|/|\Delta\Delta {\cal E}_{12}|\times |\Delta\Delta {\cal F}_{12}|/|\Delta\Delta {\cal F}_{13}| \approx 0.86$ which is close to unity. However if sites belong to  different subparts, DCA strongly underestimates epistasis with $|\Delta\Delta {\cal E}_{15}|/|\Delta\Delta {\cal E}_{12}|\times |\Delta\Delta {\cal F}_{12}|/|\Delta\Delta {\cal F}_{15}| \approx 0.33$, i.e.\ by 3 fold. In this model as well we find that long-range correlations are essentially absent (they are smaller than 1\%), despite long-range epistasis being present. Hence, a functional constraint on the cooperation between subparts potentially far away in the structure, as allosteric and active site, implies strong long-range epistasis, but does not imply strong long-range correlations, which is then reflected in small couplings. To summarize these facts, numerical values for correlation, epistasis and inferred couplings are listed in Table I. Overall, this situation is precisely that of  the \emph{in silico} allosteric material (Figs. \ref{fig:ddf_vs_J} and \ref{fig:epi_corr}), supporting that the present toy model captures the essence of the DCA limitations in more realistic settings.

\begin{table}
\begin{center}
\begin{tabular}{| c | c c c c |}
\hline
& $|\Delta\Delta{\cal F}_{ij}|$ & $C_{ij}$ & $ |\Delta\Delta{\cal E}_{ij}|$ & $J_{ij}$ \\
\hline
Same group & $1$ & $0.061$ & $0.51$ & 1.18\\  
Same subpart & $0.33$ & $-0.08$ & $0.14$ & -1.01  \\
Different subpart & $0.43$ & $0.00$ & $0.07$ & 0.40 \\
\hline
\end{tabular}
\caption{Table summarizing true and predicted epistasis magnitude, $|\Delta\Delta{\cal F}_{ij}|$ and $|\Delta\Delta{\cal E}_{ij}|$, connected correlations $C_{ij}$ and inferred couplings $J_{ij}$ in the simple model for sites $i$ and $j$ in the same group, in the same subpart and in different subparts. For $i$ and $j$ in different subparts (third row) the sizeable magnitude of epistasis is not reflected in the values of correlations, thus of the inferred couplings, in such a way that it is then underestimated by the DCA model. In section Sec.\ \ref{sec:toy_SI} in S1 Text, we derive $|\Delta\Delta{\cal F}_{ij}| = 21/49\,{\cal F}$ for $i$ and $j$ in the same group: since we do not predict the prefactor ${\cal F}$, we can fix $21/49\,{\cal F} = 1$ and other numbers in the first column follow from this choice.}
\end{center}
\label{fig:table_toy}
\end{table}

\subsubsection*{Empirical evidence}
\begin{figure}
\includegraphics[width=0.9\textwidth]{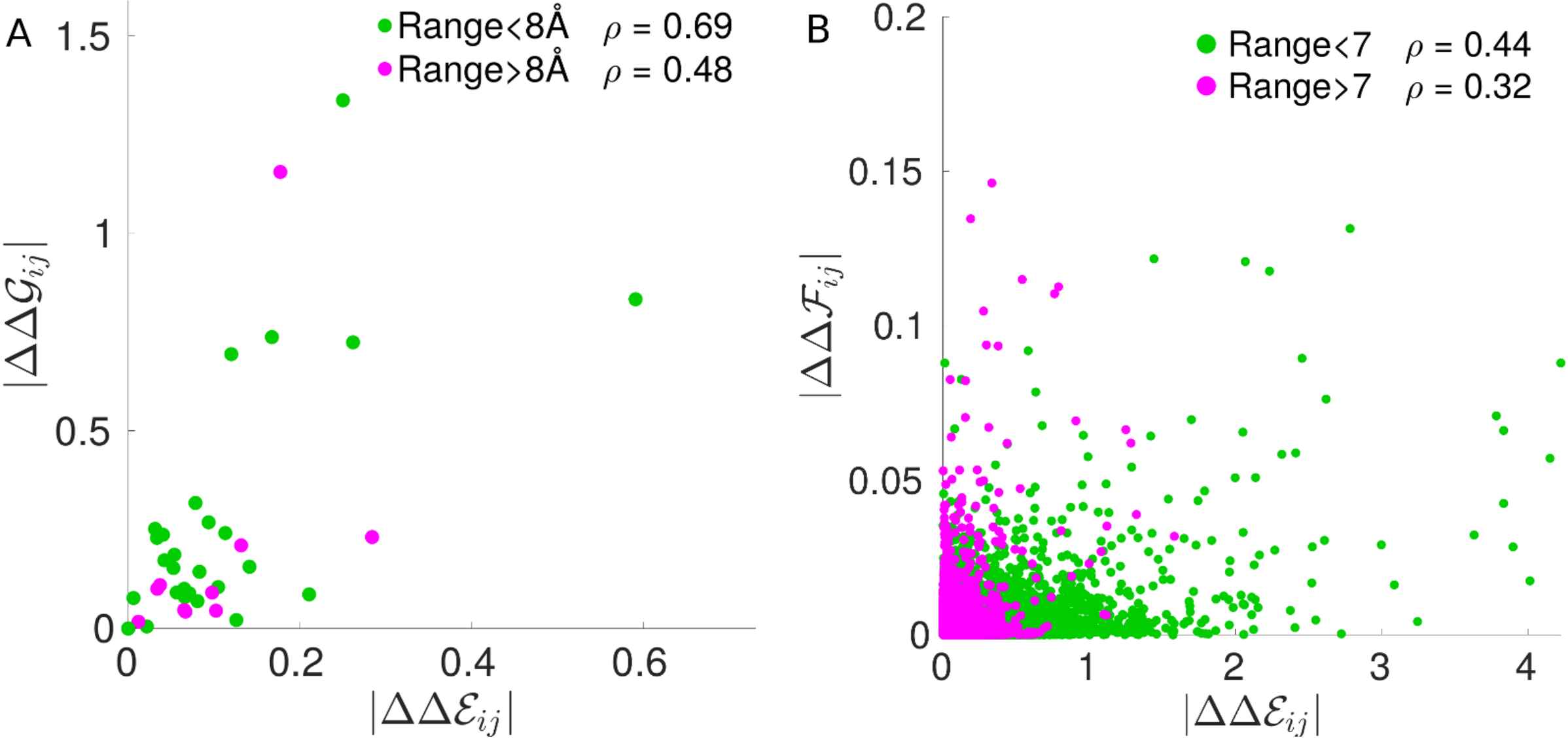}
\caption{\textbf{Prediction of experimentally measured epistasis by DCA from \cite{Salinas18}}. A: Scatter plot of average epistasis magnitude $|\Delta \Delta {\cal G}|$ vs DCA-inferred energetic couplings $|\Delta \Delta {\cal E}|$, where the color code distinguishes short and long distance pairs of residues on the PDZ $\alpha2$-helix three-dimensional structure. $\rho$, the Pearson correlation coefficient, indicates a better performance at short range. As a comparison, in B we show the scatter plot of average epistasis magnitude $|\Delta \Delta {\cal F}|$ vs DCA-inferred energetic couplings $|\Delta \Delta {\cal E}|$ in our \emph{in silico} evolved networks: similarly to A, the prediction at long distance is  poorer than at short distance.}
\label{fig:ddg_dde}
\end{figure}

Recently epistasis was measured in an empirical setting by Salinas and Ranganathan \cite{Salinas18} with the aid of deep mutational scan techniques applied to the PDZ domain $\alpha2$-helix (9 residues), which is part of an allosteric regulatory mechanism controlling ligand binding. Five homologs of PDZ domain were considered in the study. There, epistasis is 
\begin{equation}
\Delta\Delta\mathcal{G}_{ij}^{xy}=\left( \Delta \mathcal{G}_i^x+\Delta \mathcal{G}_j^y\right)-\Delta\mathcal{G}_{ij}^{xy}
\end{equation}
where $\mathcal{G}$ is the binding free energy and $x,y$ correspond to mutations happening at positions $i,j$, respectively.
DCA inference in \cite{Salinas18} was performed on an alignment of 1656 eukaryotic PDZ domains (Poole alignment, see \cite{Salinas18}), from where the DCA epistasis prediction $|\Delta \Delta {\cal E}^{xy}_{ij}|$ could be directly estimated. The authors then considered averages over mutations $x, y$ and the 5 homologs (we denote them simply as $\Delta \Delta {\cal E}_{ij}$ and $\Delta \Delta {\cal G}_{ij}$); in Fig.\ \ref{fig:ddg_dde}A we show how well $|\Delta \Delta {\cal E}_{ij}|$ predict the experimental energetic couplings $|\Delta \Delta {\cal G}_{ij}|$ for pairs of residues $(i,j)$ at distance $ > 8$\AA\ and $ < 8$\AA, where distances are measured on the known three-dimensional crystal structure of the PDZ $\alpha2$-helix and averaged over the 5 homologs. We find a stronger correlation between $|\Delta \Delta {\cal G}|$ and $|\Delta \Delta {\cal E}|$ for short range pairs (Pearson correlation $\rho=0.69$), than for long range pairs ($\rho=0.48$), as the long-range strong epistatic interaction between residues 1 and 8 is not captured by the DCA-inferred energetic couplings, see discussions in \cite{Salinas18}. $|\Delta \Delta {\cal G}_{18}|$ in Fig.\ \ref{fig:ddg_dde}A is the point at largest $|\Delta \Delta {\cal G}|$ in the long-range set. This observation is consistent with our model prediction, shown in Figs.\ \ref{fig:ddf_vs_J} and \ref{fig:ddg_dde}B, on the limits of DCA in capturing strong long-range epistasis. 

It would be important to test more broadly this predicted effect, which may be possible thanks to the advances of deep mutational scans.

\section*{Discussion}
We have benchmarked DCA in a model of protein allostery where a mechanical task must be
achieved over long distances. Such models display a rich pattern of epistasis, which can be both short and long-range and vary in sign. DCA predicts well mutation costs but is not a good generative model. This failure echoes with the drastic underestimation of long-range epistasis by the pairwise couplings inferred by DCA from evolutionary correlations. This finding rationalizes why there is no statistical evidence for long-range couplings in allosteric proteins analyzed by DCA \cite{Anishchenko2017}, where long-range epistasis and functional effects are however found \cite{olson2014,Ranghanatan2011,Salinas18}, as tested here with the data from \cite{Salinas18}.

Yet, as we show in S1 Text (see Sec.\ \ref{sec:epi_SI}), we expect that DCA can capture some aspects of the long-range epistasis pattern in allosteric proteins. Indeed, high-cost mutations exhibit stronger epistasis than low-cost ones (as also seen in RNA sequences \cite{Wilke2001,Lalic2012}, in the enzyme TEM-1 $\beta$-lactamase \cite{Krug2013} and in previous \emph{in silico} evolution work \cite{Dutta2017}), 
and are well-predicted by DCA. Specifically, the scaling of epistasis of Eq.\ \ref{eq:scalingddf} suggests as  approximation  $|\Delta\Delta {\mathcal F}_{ij}| \propto \min(\Delta {\mathcal E_i}, \Delta {\mathcal E_j})$ where $\Delta {\mathcal E}$ are inferred by DCA. Testing this prediction for epistasis patterns empirically could be made possible by the increasing availability of deep mutational scans \cite{Fowler14,Salinas18}.

Moreover, we have provided the more general argument, illustrated by a simple model, that a co-evolution based maximum-entropy approach as DCA is not the appropriate inference framework when function requires several, variable parts to work in concert. Can one find better generative models than DCA for such complex  functions? Several ways have been proposed to go beyond pairwise models by including nonlinearities, which implicitly take into account correlations at all orders, as nonlinear potentials in Restricted Boltzmann Machines \cite{tubiana2019}, maximum-entropy probability measures with a nonlinear function of the energy \cite{Humplik2017}, maximum-likelihood inference procedures based on nonlinear functions \cite{otwinowski2018} and, finally, deeper architectures \cite{riesselman2018,alley2019}. 
As a first test, we have trained a 3-layers feedforward neural network with nonlinear (sigmoid) activation functions to learn the values of fitness in the simple model of Fig.\ \ref{fig:logic_gate} and we have obtained that mutation costs and epistasis can be correctly captured by this method (see Sec.\ \ref{sec:FF_SI} in S1 Text). This observation raises the possibility that neural networks may lead to better generative models in proteins, a hypothesis that could also be benchmarked \emph{in silico}.
\\ 
Finally, as a future direction it would be interesting to extend our model by considering the constraint that the protein must fold to operate, in addition to the allosteric constraint considered here. It could be done for example in the spirit of \cite{Jacquin2016} by considering that nodes are amino-acids, and that the stiffness of the spring between two adjacent amino-acids as well as their contribution to the total folding energy depend on the identity of that pair. Although we believe that such a model will lead to similar results as presented here for long-range coupling, it will presumably differ significantly in the statistics of short range ones. In particular, it may capture  why 3-body correlations are well described by 2-body correlations in real proteins, and lead to stronger conservation overall \cite{riesselman2018}.

\section*{Methods}
\subsection*{Direct Coupling Analysis: inference procedure}
\label{sec:inference}
In a maximum-entropy approach, extracting information from MSAs can be cast as an inverse problem, i.e.\ inferring the set of parameters which enable the model (an Ising model in our setup) to reproduce certain observed statistical properties \cite{Nguyen2017,Romano17}. The exact solution of this problem is found by Maximum Likelihood algorithms, which search for the set of couplings $J_{ij}$ and fields $h_{i}$ maximizing the likelihood that the model specified by such parameters produced data with the given statistics (single-site and pairwise frequencies in our case). This exact maximization might often be infeasible, therefore to tackle the inverse problem approximate techniques have been developed: for instance, we resort to the Adaptive Cluster Expansion (ACE), an expansion of the entropy (which indeed corresponds to the likelihood) into contributions from clusters of spins \cite{Cocco11, Cocco12, Barton16}. We use the package made available by Barton 
\url{https://github.com/johnbarton/ACE}. The implementation consists of first a run of ACE followed by a proper maximum likelihood refinement (QLS routine), which takes as starting set of fields and couplings the ACE-inferred ones. Different parameters for the ACE and QLS routines can be set by the user, e.g.\ $\gamma_2$, the $L_2-$norm regularization strength for couplings which penalizes spurious large absolute values induced by undersampling and for which a natural value is $\gamma_2 =1/M$ ($M$ being the size of the sample). To help convergence, we have chosen for ACE a higher value $\gamma_2 = 10^{-2}$ and $\theta = 10^{-5}$ (this is the threshold at which the algorithm will run then exit, see \cite{Cocco12}). In the further refinement by QLS, we have set $mcb$, the number of Monte Carlo steps used to estimate the inference error, to 200000 and $\gamma_2 = 1/M$. Having full control of the numerical evolution, we have tried to avoid undersampling issues by generating a large number of configurations $M = 135000$, which leads to $\gamma_2 \approx 0.7 \times 10^{-5}$. For the inference we remove from sequences the 6 links at the active and allosteric sites as they are always associated to the symbol 1 (always occupied by a spring), so the number of parameters to infer is $ N_c'+N_c'(N_c'-1)/2 \sim 81000$ with $N_c'=N_c-6=402$. We have verified that low values of the $L_2$-regularization allow us to obtain the maximal generative performance compatible with the model (in comparison to higher regularization). By default the $L_2$ regularization of fields is $0.01\times\gamma_2$. In Fig.\ \ref{fig:DCA}A, it is shown that the result of the inference is a model perfectly able to reproduce the first and second order statistics (as it should by construction) but that fails at reproducing higher order statistics.

For a comparison, we have considered also mean field Direct Coupling Analysis (mfDCA) \cite{Morcos11}, derived from a mean-field factorized ansatz for the Boltzmann-Gibbs distribution Eq.\ \ref{eq:BG}. Couplings in mfDCA are given by $J_{ij} = -(\bm{C}^{-1})_{ij}$, where $\bm{C}_{ij} = \langle \sigma_i \sigma_j\rangle - \langle \sigma_i\rangle \langle \sigma_j\rangle$ is the covariance of the MSA (we recall that in each sequence $\sigma_i=1$ stands for the presence of a spring at link $i$ and $\sigma_i=0$ for its absence). Typically $\bm{C}$ is not invertible due to undersampling, making it necessary to add a pseudocount $\lambda$ (see \cite{Review_protein_seq}). As shown in \cite{barton2014}, a pseudocount also helps correct for the systematic biases introduced by the mean field approximation: for this reason, we have used a pseudocount $\lambda$ and chosen its value as $\lambda=0.5$, which allows the best comparison to the ACE and maximum likelihood results, see Fig.\ \ref{fig:DCA}B. It is noteworthy that in this way a computationally cheap technique as mfDCA yields a pattern of top $J_{ij}$ strikingly similar to the one of a very accurate inference achieved by the combination of ACE and maximum likelihood. Therefore mfDCA, while extremely poor as a generative model, exhibits a good performance at reconstructing the distribution of relevant couplings, as shown in Fig. \ref{fig:DCA}C. 

\subsection*{Mutation costs and generative performance in the inferred Ising model}
\label{sec:epi_inferred_app}
Costs of double mutations, i.e.\ joint mutations affecting links $i$ and $j$, can be computed in the original model via fitness changes $\Delta {\cal F}_{ij} = {\cal F} - {\cal F}_{ij}$, where ${\cal F}_{ij}$ is the fitness after springs in $i$ and $j$ have been mutated. A double mutation can correspond either to (i) adding two springs at links $i$ and $j$ (i.e.\ $\sigma_i = \sigma_j = 1$) or removing them (i.e.\ $\sigma_i = \sigma_j = 0$) or to (ii) moving a spring from link $i$ to link $j$ or viceversa (i.e.\ $\sigma_i=0$, $\sigma_j=1$ or $\sigma_i=1$, $\sigma_j=0$). Let us call the former ``non-swap'' mutations and the latter ``swap'' mutations. Swap mutations conserve the total amount of springs (360), thus the overall average coordination $\langle z\rangle = 5$, and are the ones performed in the \emph{in silico} evolution. As optimal allosteric configurations maximize fitness with respect to this type of mutations, we stick to them also when we compare mutation costs in terms of fitness and inferred energy 
(see Fig.\ \ref{fig:mutations_coop1}C): we define ``effective'' single mutation costs $\Delta {\cal F}_{i}$ and $\Delta {\cal E}_{i}$ by taking, for each link, the swap with a link in the external region (more rigid, as visible in e.g.\ Fig.\ \ref{fig:prop_gen_coop}), where mutations are completely neutral, thus whose cost would be roughly zero. 

For the generative step, we implement a Monte Carlo sampling which relocates springs from an occupied to an unoccupied link, i.e.\ which follows swap-type dynamics as for the original numerical evolution. This allows us to select, from the inferred model, sequences that are structurally as close as possible to the initial data, i.e.\ with the same average coordination $\langle z\rangle = 5$, to make a consistent comparison with them. We have verified that even relaxing this constraint in the sampling leads to sequences endowed with higher internal variability yet lying in the same range on fitness (hence the inferred model incorporates rather well the information on the fixed amount of springs). The parameters of the Ising model are inferred in such a way as to match single-site occupancy, which reflects the spatial pattern of coordination in the allosteric networks. In Fig.\ \ref{fig:prop_gen_coop} we show that generated sequences, despite having lower fitness, reproduce successfully this property as they should.

\subsubsection*{Comparison with conservation}
Single-site frequency in protein alignments, informative about local conservation, is a standard measure of mutation costs at a certain position \cite{Kumar2009} and can be fit by an independent-site Ising model. Energy (Eq.\ \ref{eq:energy}) in this case contains only field terms and, once these are inferred from link occupancies $\langle \sigma_i \rangle$, one can compute energy changes $\Delta {\cal E}_i$ upon point mutations. The energy cost of a mutation in an independent-site model is then $\Delta {\cal E}_i = (2\sigma_i - 1) h_i$, where $h_i = \log (\langle \sigma_i \rangle(1-\bar{\sigma})/\bar{\sigma}(1-\langle \sigma_i \rangle))$ describes how the observed occupancy of a link $i$, $\langle \sigma_i \rangle$, is biased away from the average occupancy $\bar{\sigma}=360/408=0.88$. In average $\Delta {\cal E}_i$ gives also a measure of \emph{conservation} of link $i$ as it is 0 when $\langle \sigma_i \rangle = \bar{\sigma}$ and it increases the more link $i$ tends to be either occupied or vacant. The improvement achieved by the pairwise model over this conservation-based measure of mutation costs is extremely significant (see inset of Fig.\ \ref{fig:mutations_coop1}C). On the one hand, conservation is a purely local measure - it takes into account how a particular position is crucial to the propagation of the allosteric response. Including pairwise couplings proves to be crucial to capture the context-dependence of mutation costs, and thus must be included for their quantitative prediction. On the other hand, the degree itself of structural conservation is rather low due to the heterogeneity of the shear-design MSA: the conformation, precise location and size of the shear path, hence the role of each link, can vary from architecture to architecture, leading to low structural conservation (with peaks only around the active and allosteric site). Conservation is found much higher \emph{within} one set of dynamically related solutions (as for Fig.\ \ref{fig:epistasis}A), corresponding to one realization of the shear design among the many included in the MSA (see in particular Fig.\ 4G in \cite{Yan18}).

\subsubsection*{\bf Acknowledgment:} 
We acknowledge interesting and stimulating discussions with Eric Aurell, John Barton, Johannes Berg, Simona Cocco, Paolo de Los Rios, 
Solange Flatt, Joachim Krug, Michael Lassig, Duccio Malinverni, Simone Pompei, Remi Monasson, Martin Weigt, Le Yan, Stefano Zamuner. 
We are particularly grateful to John Barton, Le Yan, Duccio Malinverni and Stefano Zamuner for help with the codes and to Rama Ranganathan for making available data from \cite{Salinas18}.

\bibliography{Wyartbibnew_Barbara}

\noindent\fcolorbox{white}{white}{%
    \parbox{0.9\textwidth}{%
    {\bf Supporting Information legends} \\
S1 Text. Supporting information for Direct Coupling Analysis of Epistasis in Allosteric Materials.\\
}
}

\include{bioarxiv_SI}

\bibliographystyle{unsrt}
\bibliography{vancouver}
% Bibliography

\end{document}

%% file: bioarxiv_SI.tex
\setcounter{page}{1}
\pagenumbering{arabic} 

\title{\begin{center}\LARGE{Supplementary Information S1 Text: \vspace{0.5cm} \\
Direct Coupling Analysis of Epistasis in Allosteric Materials}\end{center}}
\vspace{0.5cm}
\begin{center}Barbara Bravi$^1$, Riccardo Ravasio$^1$, Carolina Brito$^2$, Matthieu Wyart$^1$ \end{center}

\vspace{0.5cm}
%\begin{flushleft}
\begin{center}
\emph{$^1$ Institute of Physics, \'Ecole Polytechnique F\'ed\'erale de Lausanne, CH-1015 Lausanne, Switzerland}\\
\emph{$^2$ Instituto de F\`isica, Universidade Federal do Rio Grande do Sul, CP 15051, 91501-970 Porto Alegre RS, Brazil}
%\end{flushleft}
\end{center}

\beginsupplement
\begin{figure}
\text{\myfont{A}}\\
 \includegraphics[width=0.30\textwidth]{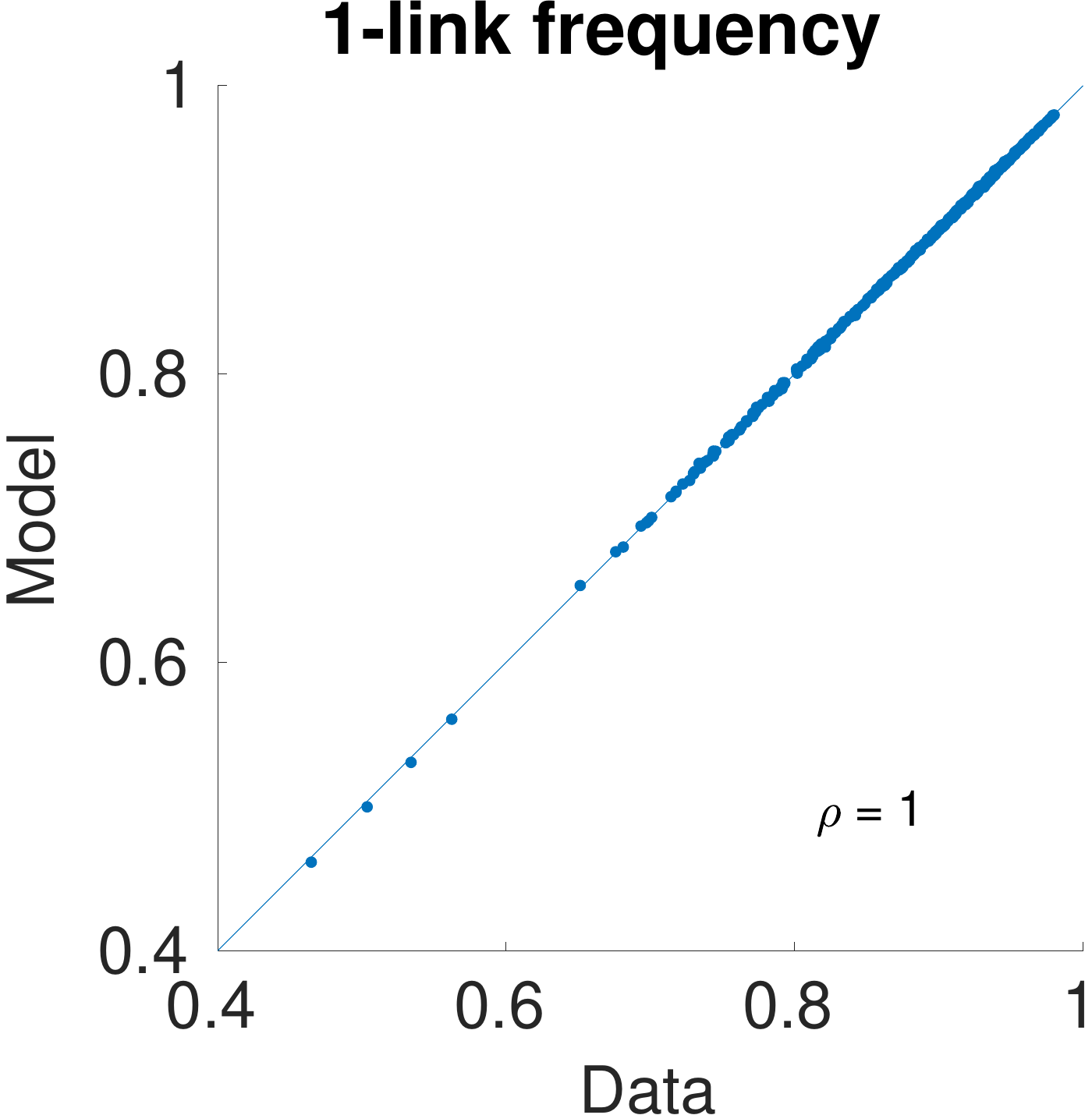}
 \includegraphics[width=0.32\textwidth]{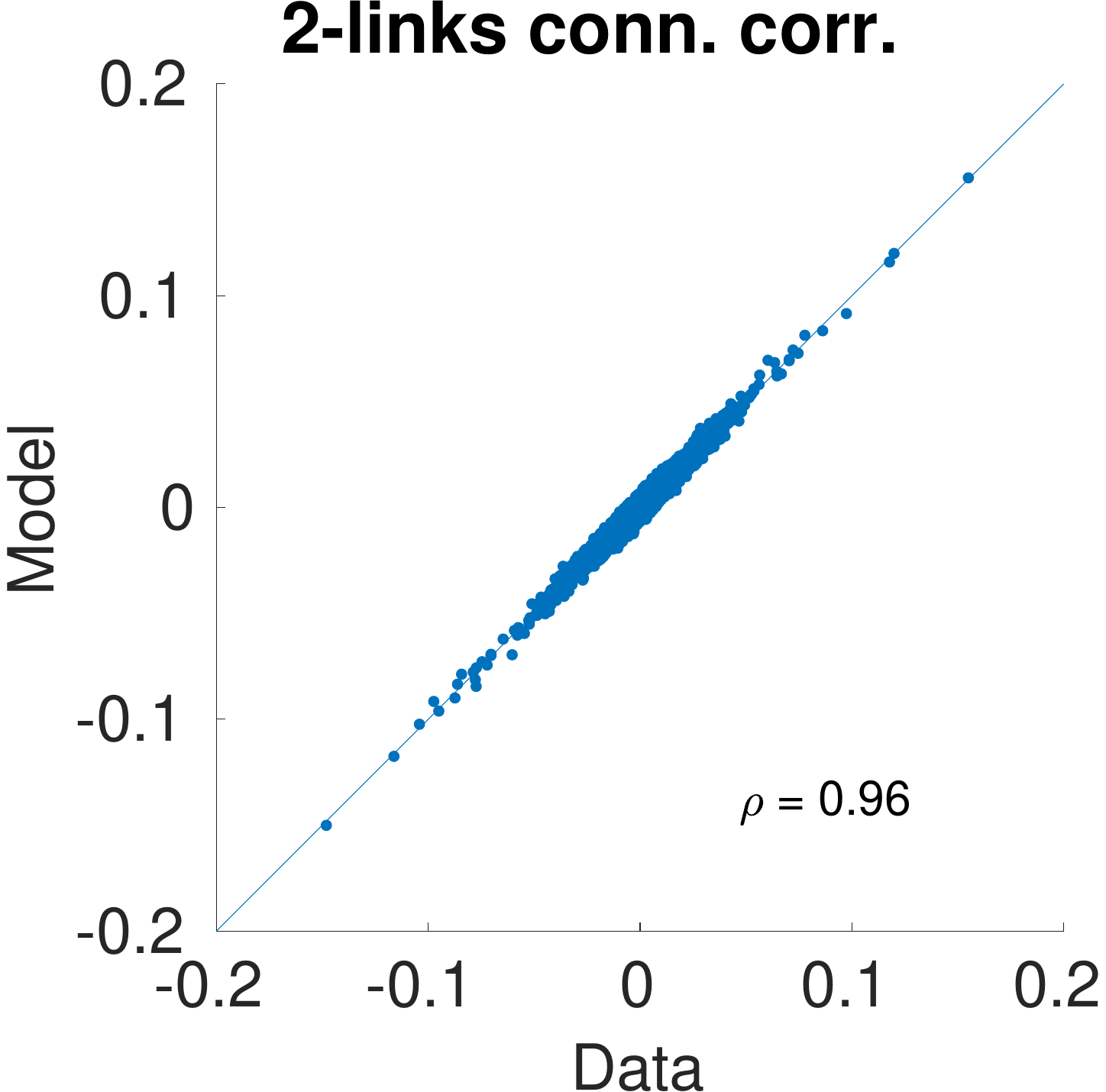}
 \includegraphics[width=0.36\textwidth]{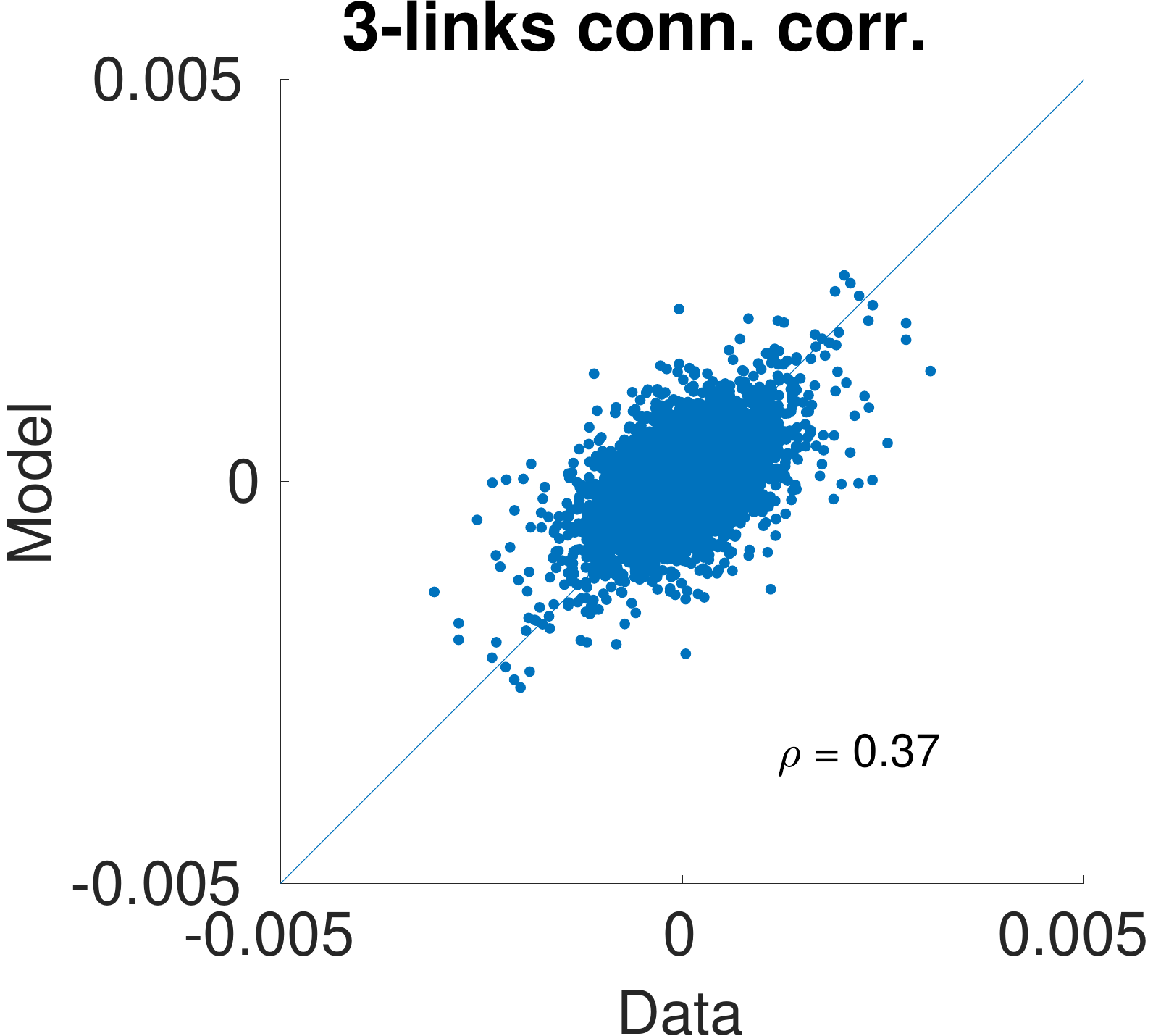}
 \vspace{2cm}\\
\hspace{1cm}\text{\myfont{B}}\hspace{8cm}\text{\myfont{C}}
\begin{center}
\includegraphics[width=0.4\textwidth]{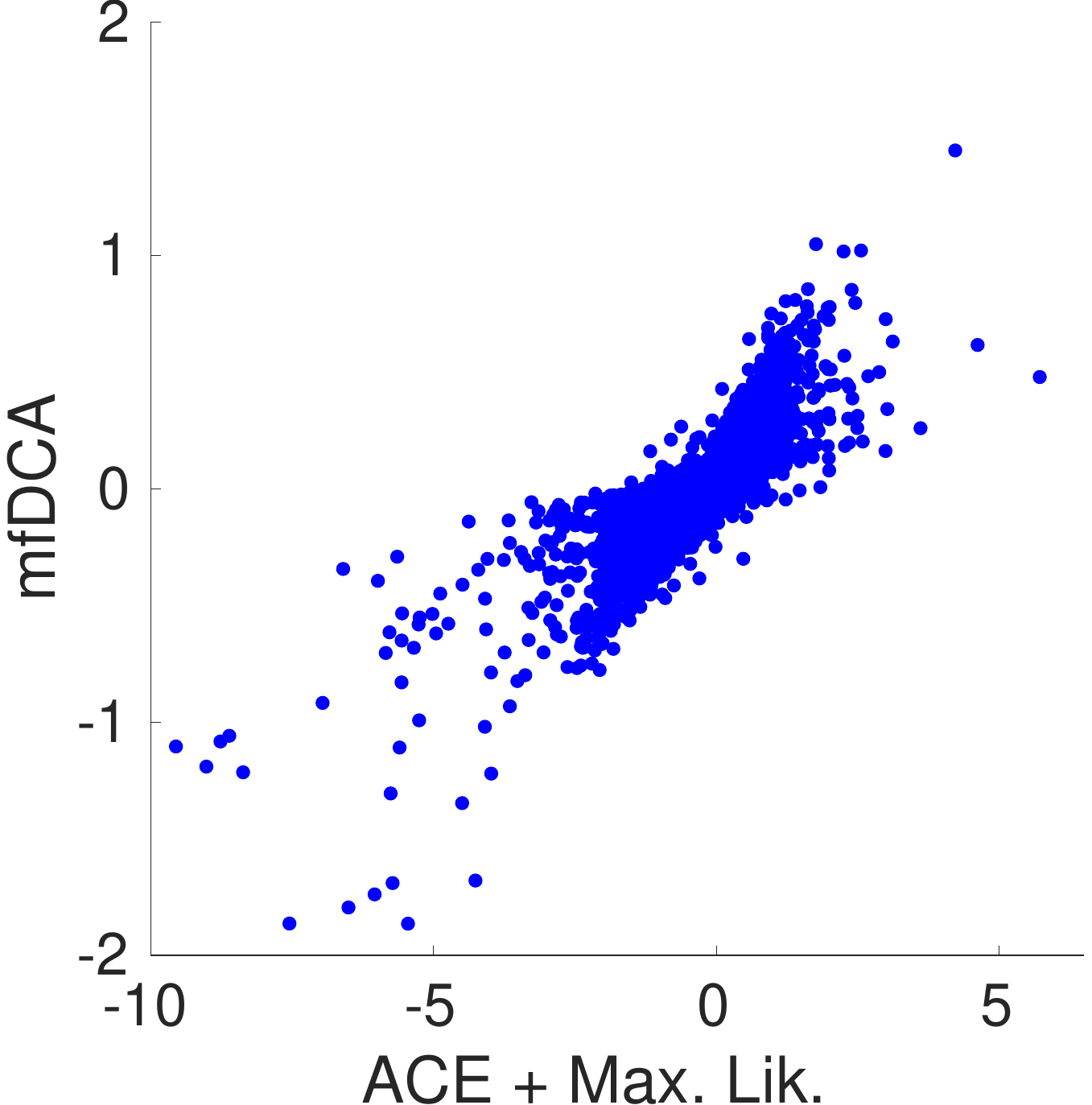}
\hspace{2cm} \includegraphics[width=0.45\textwidth]{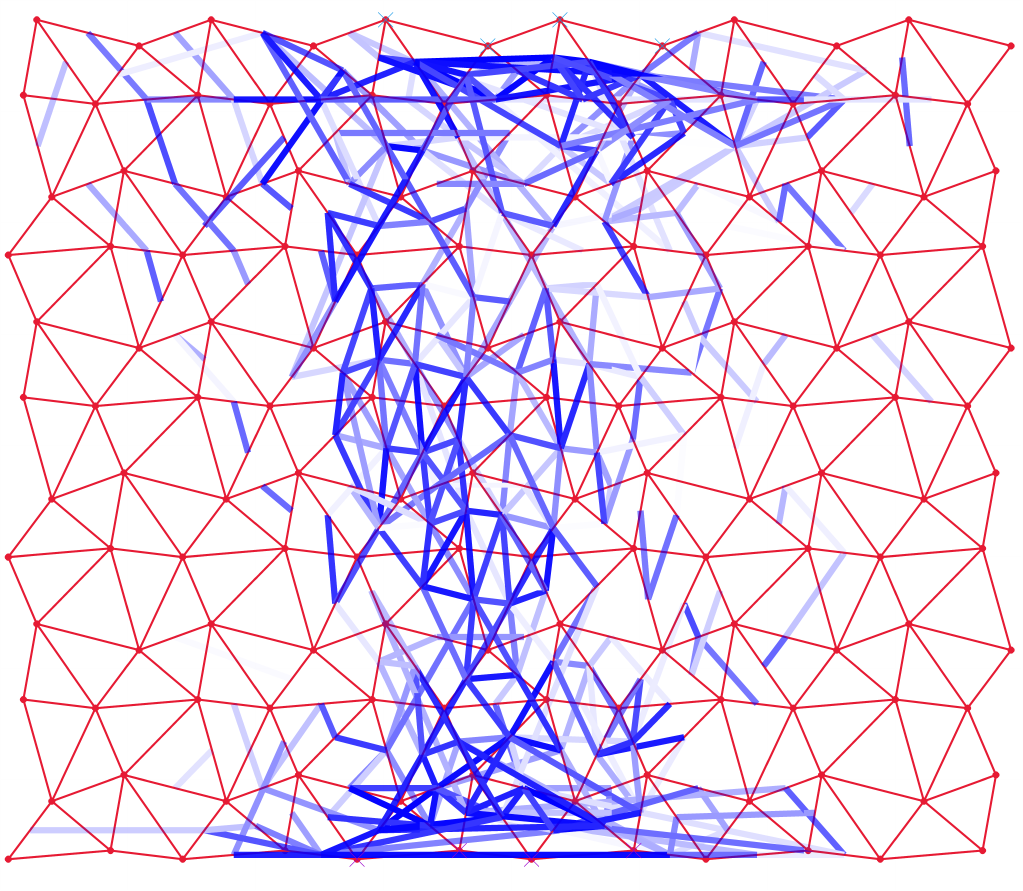}
\end{center}
\caption{\textbf{Performance of the inference procedure}. A: Statistics of the model inferred by combining ACE and Maximum Likelihood. 1-link frequency and 2-links connected correlations are very accurately reproduced, as they should by construction (the relative errors, defined as in \cite{Cocco12}, are respectively $\epsilon_m = 2.45 \times 10^{-1}$ and $\epsilon_C = 1.30\times 10^{-1}$). In contrast the third order connected correlations, which are not constrained in the inference, are not well captured (Pearson correlation coefficient $\rho = 0.37$). This is a hint that the Ising model - a pairwise probabilistic model over $\sigma_i$ - is an approximation which becomes poor for estimating higher order moments. B: Scatter plot comparing $J_{ij}$ inferred via mfDCA to the direct couplings of ACE + Max. Lik.: the pseudocount in mfDCA has been set to $\lambda=0.5$ in such a way as to obtain the highest correlation between the two. C: Spatial distribution of top 400 mfDCA-inferred couplings on the network. The reconstruction of the topology of relevant couplings is rather robust with respect to the choice of more approximate inference methods as mfDCA. As in Fig.\ \ref{fig:ddf_vs_J}A (inset) of the main text, they are concentrated at short range, i.e.\ they connect links lying close either to the active site or the allosteric site and in the central high-shear path. Long range mfDCA couplings, connecting links around respectively allosteric and active site, are weaker and appear among the top 600-1000 ones, implying an even worse performance at predicting long range epistasis than ACE + Max. Lik.}
\label{fig:DCA}
\end{figure}

\begin{figure}
\hspace{0.5cm}\text{\myfont{A}}\hspace{7.8cm}\text{\myfont{B}}
\begin{center}
 \includegraphics[width=0.4\textwidth]{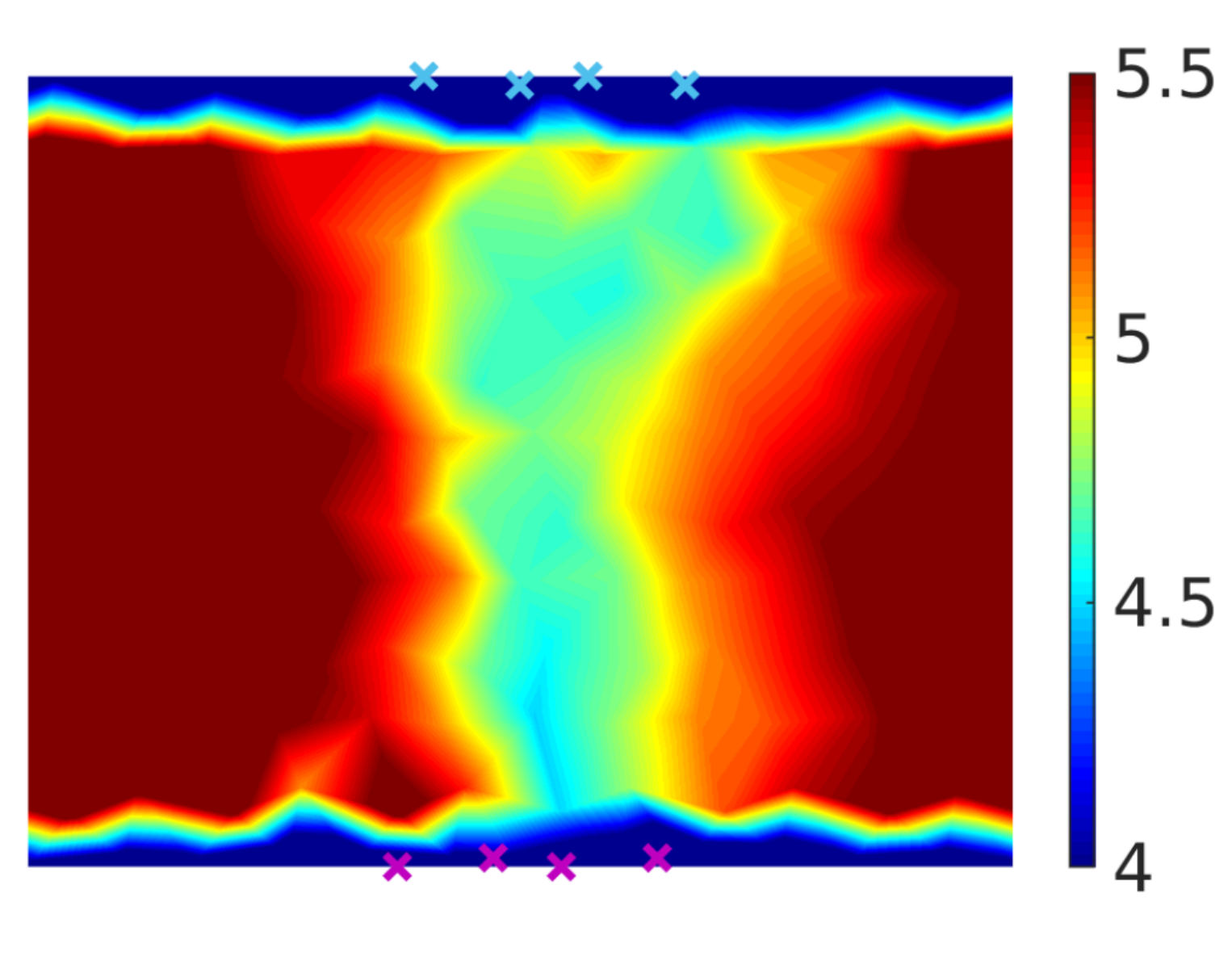}
 \hspace{1cm}
 \includegraphics[width=0.4\textwidth]{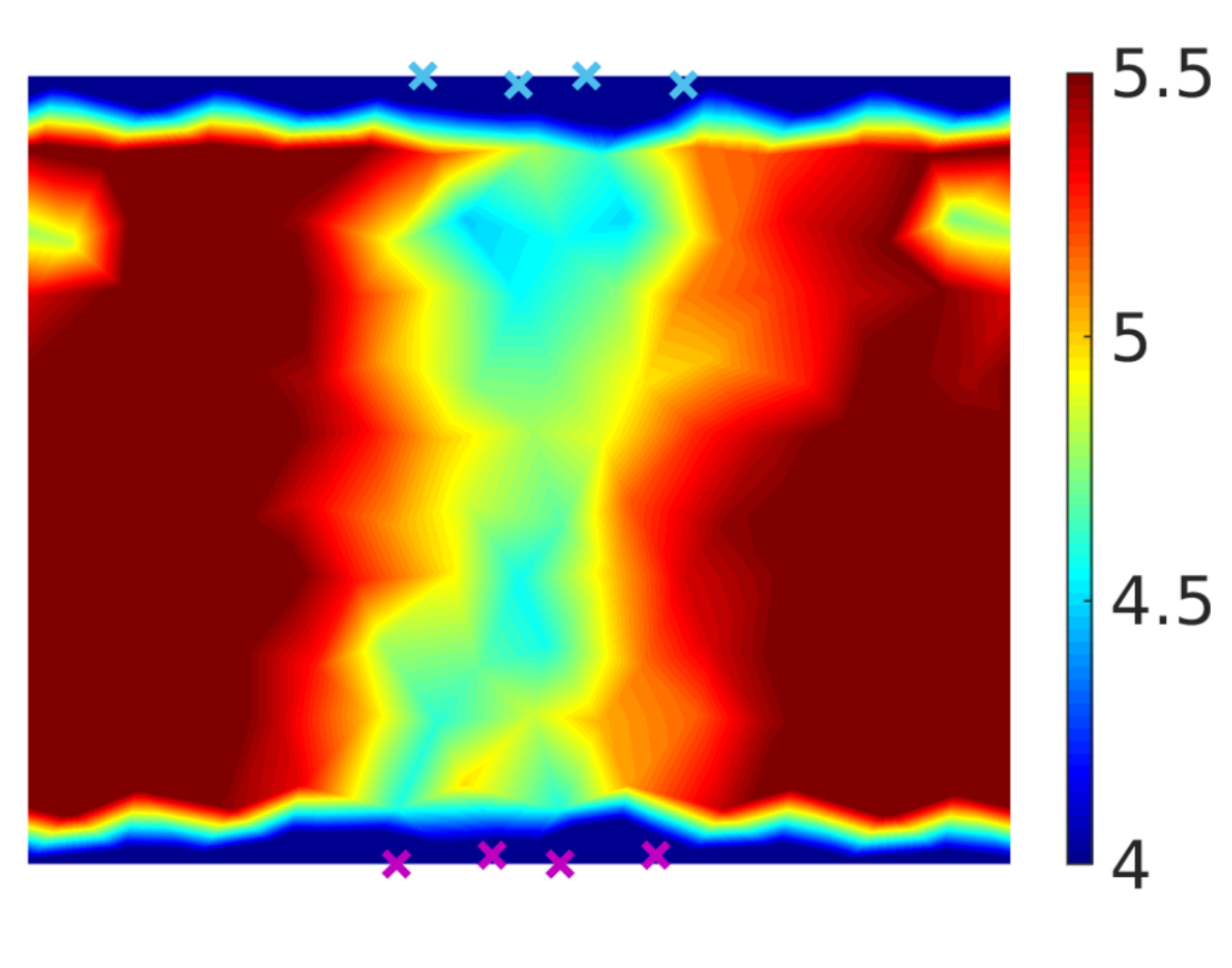}
 \end{center}
 \caption{\textbf{Properties of generated allosteric sequences}. Coordination map of original sequences (A) and generated ones (B). 
 They both exhibit a softer (i.e.\ with coordination $z<5$) central path joining active and allosteric sites (indicated respectively by blue and purple crosses) along which the shear-like sliding takes place. This path is embedded in a more connected, ``rigid'' region where the coordination $z>5$. Solutions sampled from the inferred energy landscape have the expected design but are not maximally fit, showing that more ``structural'' components, as the distribution of links, are captured but additional information would be needed to reproduce a complex mechanical function as the cooperative fitness.}
 \label{fig:prop_gen_coop}
\end{figure}

\cleardoublepage

\section{Mechanical interpretation of mutation costs and epistasis}
\label{sec:mech_epi_app}
Let us denote by $\epsilon$ the set of nodes where ligand binding takes place, e.g.\ for ligand binding at the allosteric site $\epsilon = ({\cal A}l)$ with size $\dim(\epsilon)=n_0$. Such event imposes a displacement ${\bf R}^{\epsilon}$ on the nodes $\epsilon$ which imparts locally a force $\bm{F}^{\epsilon}$ and induces a response ${\bf R}^{\epsilon \rightarrow r}$ on all the other nodes $r$. Clearly $\dim(\epsilon) + \dim(r)=L^d$ where $L^d$ is the total number of nodes for a network of size $L$ in $d$ dimensions; for the example of binding to the allosteric site $r=({\cal A}c, b)$, where $b$ stands for the ``bulk'' of nodes belonging neither to the allosteric nor to the active site. (In this paper we consider networks as in Fig.\ \ref{fig:msa}A of the main text, with $d=2$, $L=12$ and $n_0=4$ for both active and allosteric site). Considering  the deformation as a linear response to the external force, the relation between force and overall response field is written in terms of the dynamical matrix $\bm{{\cal M}}$
\be
\label{eq:mforce}
\lp\begin{array}{c}
\bm{F}^{\epsilon}\\
\bm{\mathbb{0}}
\end{array}\rp
=\bm{{\cal M}}
\lp\begin{array}{c}
{\bf R}^{\epsilon}\\
{\bf R}^{\epsilon \rightarrow r}
\end{array}\rp
\ee
hence $\bm{{\cal M}}$ is endowed with a block structure as follows
\[\bm{{\cal M}}=\begin{pmatrix}
\bm{{\cal M}}^{\epsilon, \epsilon} & \bm{{\cal M}}^{\epsilon, r} \\
(\bm{{\cal M}}^{\epsilon, r})^{T}   & \bm{{\cal M}}^{r, r}  \\
\end{pmatrix}\]
For pairwise interactions such as springs, $\bm{{\cal M}}$ is symmetric. Forces as well as responses can be calculated solely from the imposed displacement by introducing a matrix $\bm{{\cal Q}}$
\[\bm{{\cal Q}}=\begin{pmatrix}
\bm{\mathbb{1}}^{\epsilon} & -\bm{{\cal M}}^{\epsilon, r} \\
\bm{\mathbb{0}}   & -\bm{{\cal M}}^{r, r}  \\
\end{pmatrix}\]
such that
\be
\label{eq:qforce}
\lp\begin{array}{c}
\bm{F}^{\epsilon}\\
{\bf R}^{\epsilon \rightarrow r}
\end{array}\rp
=\bm{{\cal Q}}^{-1}\bm{{\cal M}}
\lp\begin{array}{c}
{\bf R}^{r}\\
\bm{\mathbb{0}}
\end{array}\rp
\ee
Binding at ${\epsilon}$ costs an elastic energy $E^{\epsilon}$ 
\be 
\label{eq:enep}
E^{\epsilon} = \frac{1}{2}\,\bm{F}^{\epsilon}\cdot \bm{R}^{\epsilon}
\ee
and the cooperative fitness is defined by a combination of such elastic energies 
\be 
\label{eq:fitcoop}
{\cal F} =  E^{{\cal A}c} - (E^{{\cal A}c,{\cal A}l} - E^{{\cal A}l})
\ee
where $E^{{\cal A}c}$, $E^{{\cal A}c,{\cal A}l}$ and $E^{{\cal A}l}$ are given by Eq.\ \ref{eq:enep} with $\epsilon = ({\cal A}c)$, $\epsilon = ({\cal A}c,{\cal A}l)$ and $\epsilon = ({\cal A}l)$ respectively. Maximal cooperativity corresponds to making binding of a substrate at the active site energetically favored when already a ligand is bound to the allosteric site, as this reduces its binding energy from $E^{{\cal A}c}$ to $(E^{{\cal A}c,{\cal A}l} - E^{{\cal A}l})$. One can express the energy of joint binding at the allosteric and active site $E^{{\cal A}c,{\cal A}l}=\frac{1}{2}\bm{F}^{{\cal A}c,{\cal A}l}\cdot \bm{R}^{{\cal A}c,{\cal A}l}$ as 
\be 
\label{eq:approx}
\frac{1}{2}\,\bm{F}^{{\cal A}c,{\cal A}l}\cdot \bm{R}^{{\cal A}c,{\cal A}l} =
\frac{1}{2}\,\bm{F}^{{\cal A}l}\cdot \bm{R}^{{\cal A}l} + \frac{1}{2}\bm{F}^{{\cal A}c}_{|{\cal A}l}\cdot (\bm{R}^{{\cal A}c} -\bm{R}^{{\cal A}l \rightarrow {\cal A}c})
\ee
i.e.\, after binding at the allosteric site with an energy cost $\frac{1}{2}\,\bm{F}^{{\cal A}l}\cdot \bm{R}^{{\cal A}l}$, the elastic energy of binding at the active site is determined by (i) the force there when a ligand is already bound at the allosteric site ($\bm{F}^{{\cal A}c}_{|{\cal A}l}$ with subindex $|{\cal A}l$); (ii) the displacement imposed at the active site $\bm{R}^{{\cal A}c}$ to which we subtract the response already caused by ligand binding at the allosteric site $\bm{R}^{{\cal A}l \rightarrow {\cal A}c}$. Eq.\ \ref{eq:approx} allows us to rewrite Eq.\ \ref{eq:fitcoop} as
\be 
\label{eq:approxfit}
{\cal F} = \frac{1}{2}\,\bm{F}^{{\cal A}c}_{|{\cal A}l}\cdot \bm{R}^{{\cal A}l \rightarrow {\cal A}c} + \frac{1}{2}\,\delta\bm{F}^{{\cal A}l \rightarrow {\cal A}c} \cdot \bm{R}^{{\cal A}c}
\ee
where one has $\bm{F}^{{\cal A}c} - \bm{F}^{{\cal A}c}_{|{\cal A}l} = \delta\bm{F}^{{\cal A}l \rightarrow {\cal A}c}$.

We now consider the weak elastic coupling limit between the allosteric and active sites. Physically, we assume that the response induced at the active site by binding at the allosteric site is small compared to the one induced by binding at the active site. Mathematically, it corresponds to the assumptions that the elements $\bm{{\cal M}}^{{\cal A}c, b}(\bm{{\cal M}}^{b,b})^{-1}\bm{{\cal M}}^{b, {\cal A}l}$ are small. In this limit,  expressing $\delta\bm{F}^{{\cal A}l \rightarrow {\cal A}c}$ and $\bm{R}^{{\cal A}l \rightarrow {\cal A}c}$ in terms of the imposed displacements by using Eq.\ \ref{eq:qforce}, we find that each term in Eq.\ \ref{eq:approxfit} follows, at first order in the $\bm{{\cal M}}^{{\cal A}c, b}(\bm{{\cal M}}^{b,b})^{-1}\bm{{\cal M}}^{b, {\cal A}l}$: 
\be
\frac{1}{2}\,\bm{F}^{{\cal A}c}_{|{\cal A}l}\cdot \bm{R}^{{\cal A}l \rightarrow {\cal A}c} \approx
\frac{1}{2}\,\delta\bm{F}^{{\cal A}l \rightarrow {\cal A}c} \cdot \bm{R}^{{\cal A}c} \approx \frac{1}{2}\,(\bm{R}^{{\cal A}c})^T \cdot (\bm{{\cal M}}^{{\cal A}c, b})(\bm{{\cal M}}^{b,b})^{-1}(\bm{{\cal M}}^{b, {\cal A}l})\cdot\bm{R}^{{\cal A}l}
\ee
where a sum over $b$, the ensemble of ``bulk'' nodes, is taken. Hence, by using that $\frac{1}{2}\,\delta\bm{F}^{{\cal A}l \rightarrow {\cal A}c} \cdot \bm{R}^{{\cal A}c} \approx \frac{1}{2}\,\bm{F}^{{\cal A}c}_{|{\cal A}l}\cdot \bm{R}^{{\cal A}l \rightarrow {\cal A}c}$, we obtain from Eq.\ \ref{eq:approxfit}
\be
{\cal F} \approx {\bf F}^{{\cal A}c}\cdot\bm{R}^{{\cal A}l \rightarrow {\cal A}c}
\ee
since $\bm{F}^{{\cal A}c}_{|{\cal A}l}$ can be approximated by $\bm{F}^{{\cal A}c}$ in the weak coupling limit. 

\begin{figure}
\hspace{6cm}\text{\myfont{A}}
\begin{center}
\includegraphics[width=0.5\textwidth]{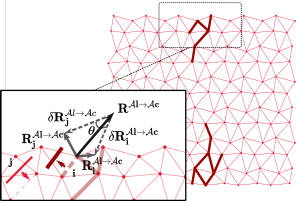}\\
\end{center}
\vspace{1cm}
\hspace{0.5cm} \text{\myfont{B}}\hspace{8.5cm}\text{\myfont{C}}
\begin{center} 
\includegraphics[width=0.4\textwidth]{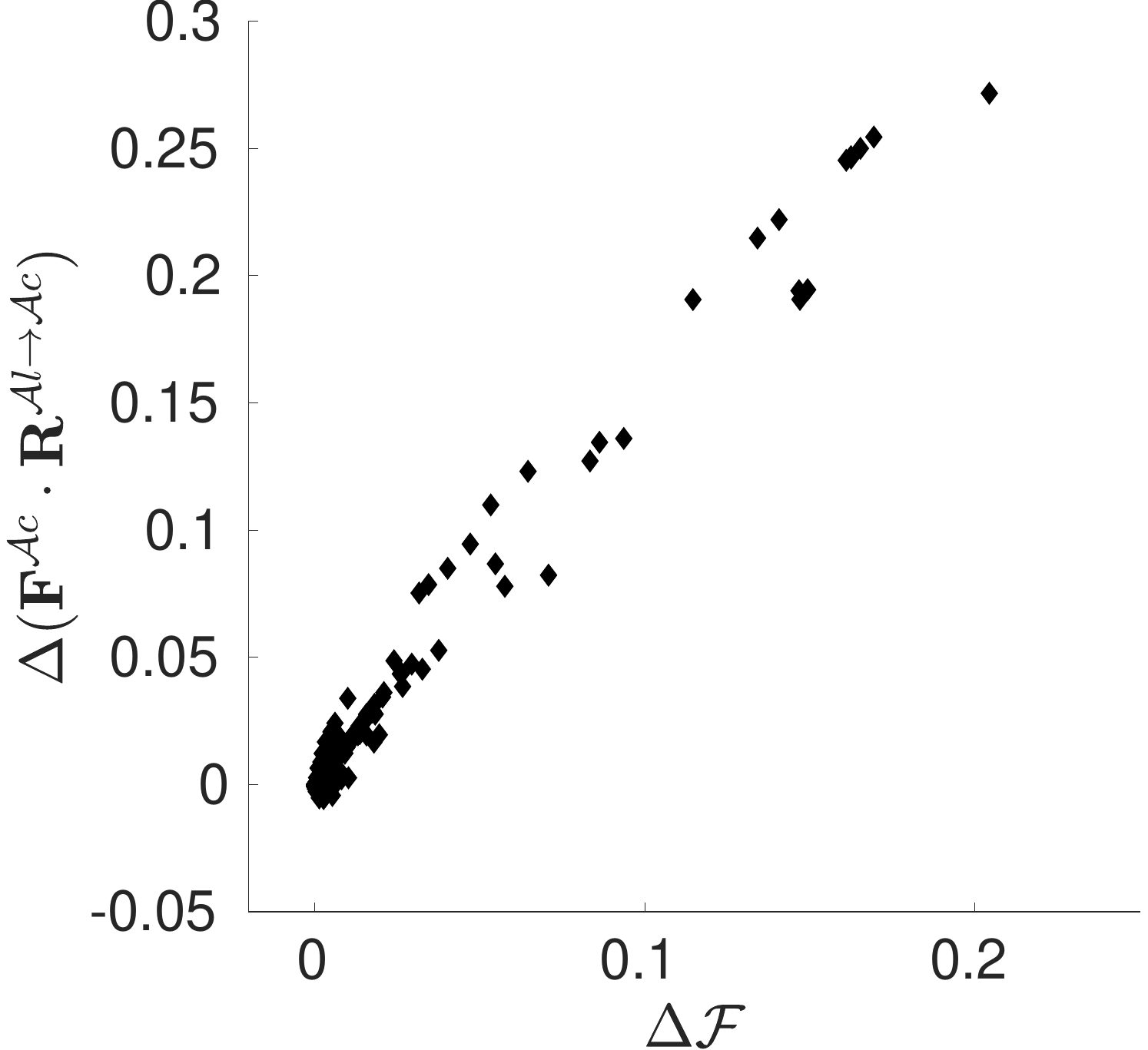}
\hspace{2cm}
\includegraphics[width=0.4\textwidth]{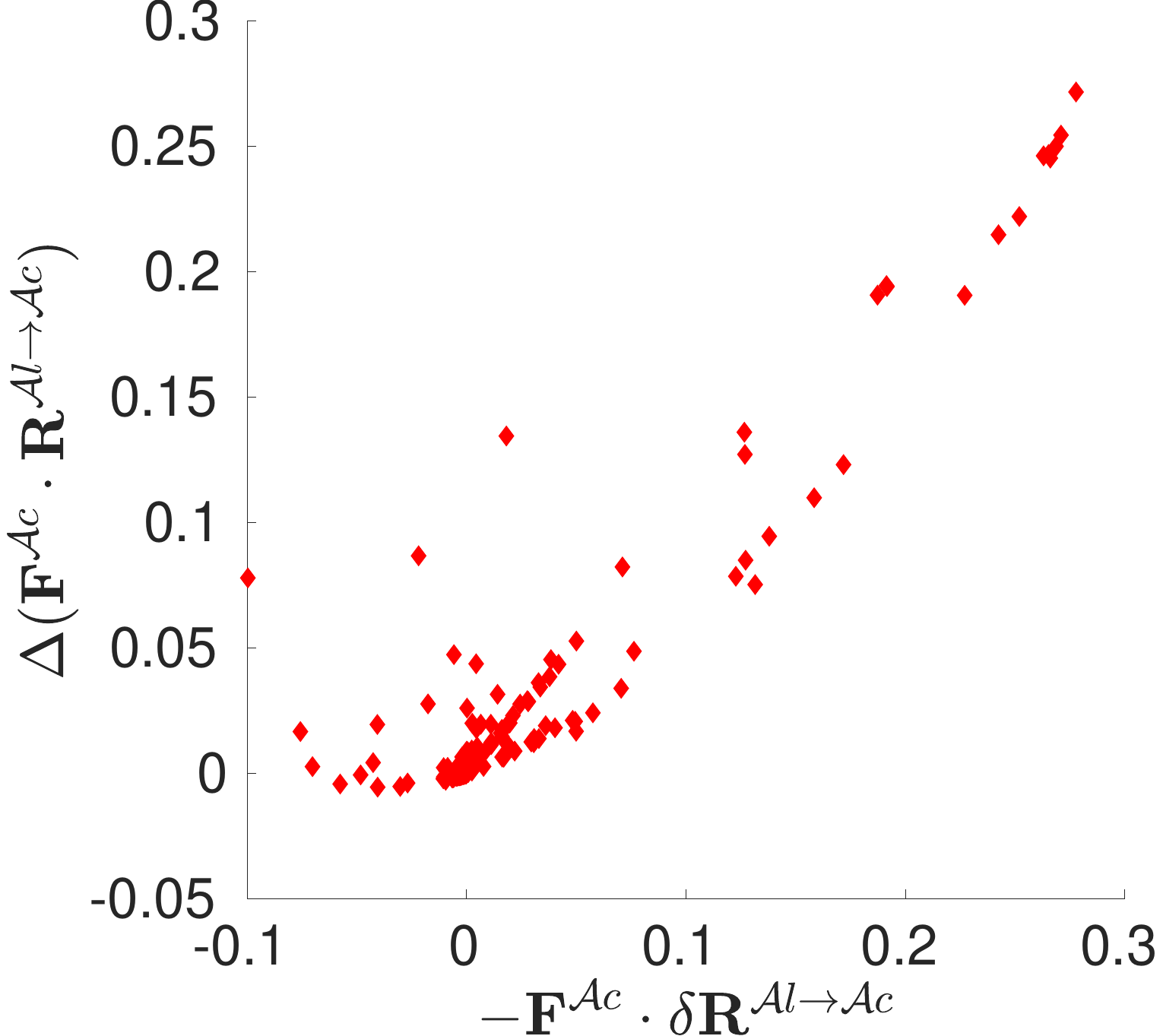}
\end{center}
\caption{\textbf{Mechanics of mutations}. A: The geometry of mutation costs is illustrated in the zoom on the active site region (note that for simplicity of visualization we consider only one of the $n_0=4$ nodes). Thick, dark red lines highlight links whose disruption would be lethal for the allosteric fitness. These few links, crucial to the long-distance propagation of the allosteric response, are located around active and allosteric site and exhibit maximal epistasis along with maximal single mutation costs (i.e.\ they populate the saturation region of Fig.\ \ref{fig:epistasis}A in the main text). After a lethal mutation consisting in removing a spring at link $i$, the displacement at the active site ${\bf R}^{{\cal A}l \rightarrow{\cal A}c}_i$ is significantly reduced with respect to the original optimal displacement ${\bf R}^{{\cal A}l \rightarrow{\cal A}c}$ and their difference is given by $\delta {\bf R}^{{\cal A}l \rightarrow {\cal A}c}_i$ (dashed arrow). When a second lethal mutation at $j$ occurs, we denote by $\theta$ the angle between $\delta {\bf R}^{{\cal A}l \rightarrow {\cal A}c}_i$ and $\delta {\bf R}^{{\cal A}l \rightarrow {\cal A}c}_j$; for lethal mutations $\cos(\theta)\approx 1$ (see Fig.\ \ref{fig:epistasis}B in the main text), i.e.\ they all tend to have a homogeneous direction of action which is precisely the one opposite to the displacement at the active site. B and C: Numerical test of the approximation $\Delta {\cal F}_i \approx \Delta({\bf F}^{{\cal A}c}\cdot\bm{R}^{{\cal A}l \rightarrow {\cal A}c})_i$ (B) and of $\Delta({\bf F}^{{\cal A}c}\cdot\bm{R}^{{\cal A}l \rightarrow {\cal A}c})_i \approx - {\bf F}^{{\cal A}c} \cdot \delta \bm{R}^{{\cal A}l \rightarrow {\cal A}c}_i$ (C). The latter is valid only for medium-high mutation costs.}
\label{fig:scatter_mech}
\end{figure}

If we denote by ${\bf F}^{{\cal A}c}_i$ and $\bm{R}^{{\cal A}l\rightarrow {\cal A}c}_i$ forces and displacements after a mutation at link $i$, the cost of one mutation can be expressed in this approximation (see Panel B Fig.\ \ref{fig:scatter_mech} for a numerical validation of our approximation) as $\Delta {\cal F}_i \approx \Delta({\bf F}^{{\cal A}c}\cdot\bm{R}^{{\cal A}l \rightarrow {\cal A}c})_i$, where $\Delta({\bf F}^{{\cal A}c}\cdot\bm{R}^{{\cal A}l \rightarrow {\cal A}c})_i = {\bf F}^{{\cal A}c}\cdot\bm{R}^{{\cal A}l \rightarrow {\cal A}c} - {\bf F}^{{\cal A}c}_i \cdot \bm{R}^{{\cal A}l\rightarrow {\cal A}c}_i$. It can be further rewritten as
\be
\Delta({\bf F}^{{\cal A}c}\cdot\bm{R}^{{\cal A}l \rightarrow {\cal A}c})_i \approx -\big({\bf F}^{{\cal A}c}\cdot \delta \bm{R}_i^{{\cal A}l \rightarrow {\cal A}c} + \delta{\bf F}_i^{{\cal A}c} \cdot \bm{R}^{{\cal A}l \rightarrow {\cal A}c} + \delta{\bf F}_i^{{\cal A}c} \cdot \delta \bm{R}_i^{{\cal A}l \rightarrow {\cal A}c} \big)
\ee
having defined changes in force as $\delta{\bf F}^{{\cal A}c}_i = {\bf F}^{{\cal A}c}_i - {\bf F}^{{\cal A}c}$ in analogy to changes in displacement $\delta\bm{R}^{{\cal A}l \rightarrow {\cal A}c}_i$ introduced in the main text. We find numerically that the cost of single mutations, when it is not too small, is dominated by the changes in displacement at the active site
\be
\label{eq:approx2}
\Delta {\cal F}_i \approx  - {\bf F}^{{\cal A}c} \cdot \delta \bm{R}^{{\cal A}l \rightarrow {\cal A}c}_i
\ee
as implied jointly by in Panels B and C Fig.\ \ref{fig:scatter_mech}. As a consequence, epistasis between mutations at $i$ and $j$ with significant magnitude can be written $\Delta \Delta {\cal F}_{ij} \approx  - {\bf F}^{{\cal A}c} \cdot \big(\delta \bm{R}_{ij}^{{\cal A}l \rightarrow {\cal A}c} - \delta \bm{R}_{i}^{{\cal A}l \rightarrow {\cal A}c} - \delta \bm{R}_{j}^{{\cal A}l \rightarrow {\cal A}c}\big)$, as presented in the main text. Displacement vectors and their changes upon high-cost mutations at the active site are schematically depicted in Panel A Fig. \ref{fig:scatter_mech}.  

\cleardoublepage
\section{Prediction of epistasis}
\label{sec:epi_SI}
\begin{figure}
\hspace{4.5cm}\text{\myfont{A}}\\
\begin{center}
\includegraphics[width=0.48\textwidth]{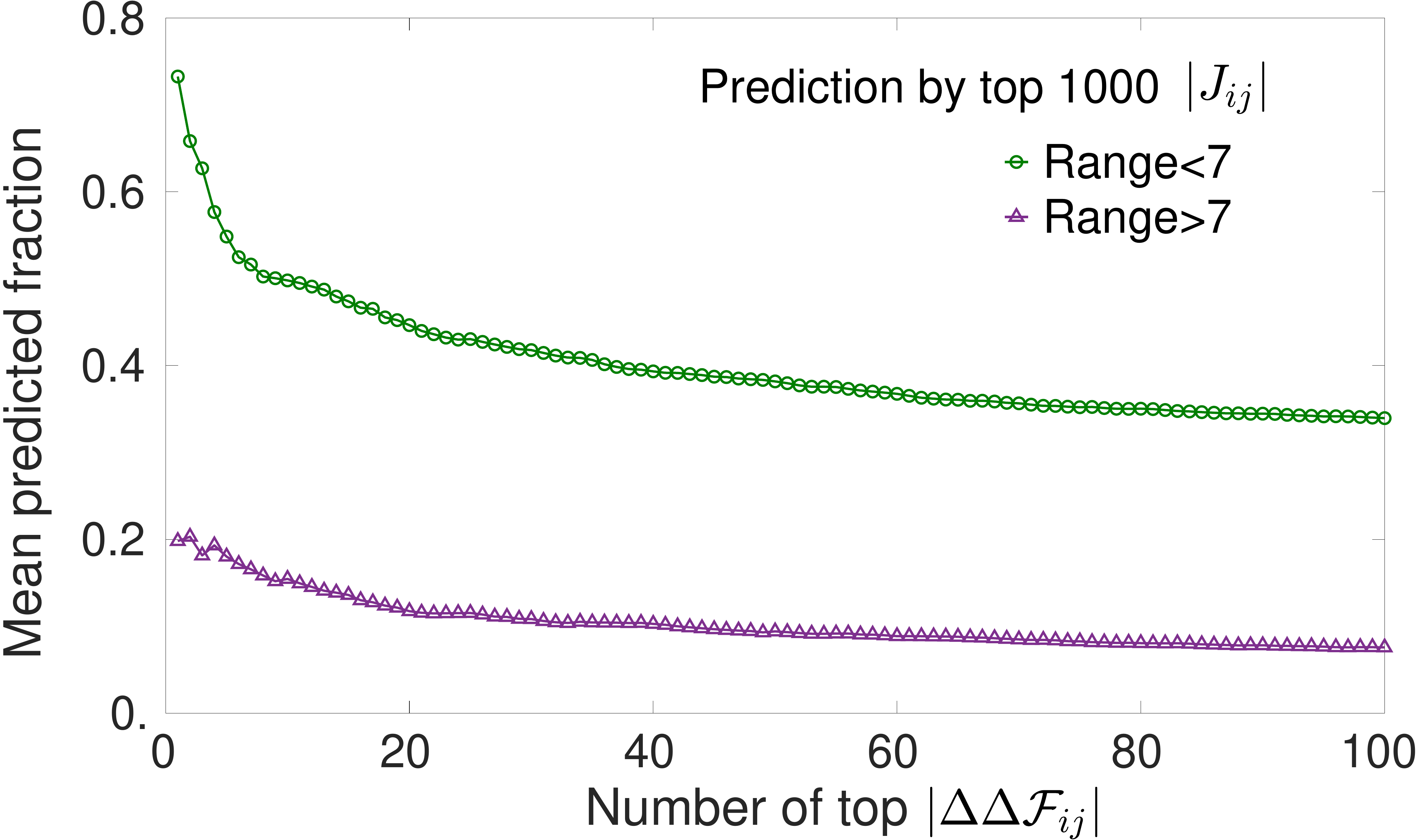}\\
\end{center}
 \flushleft
 \text{\myfont{B}}\hspace{8cm}\text{\myfont{C}}\\
 \includegraphics[width=0.48\textwidth]{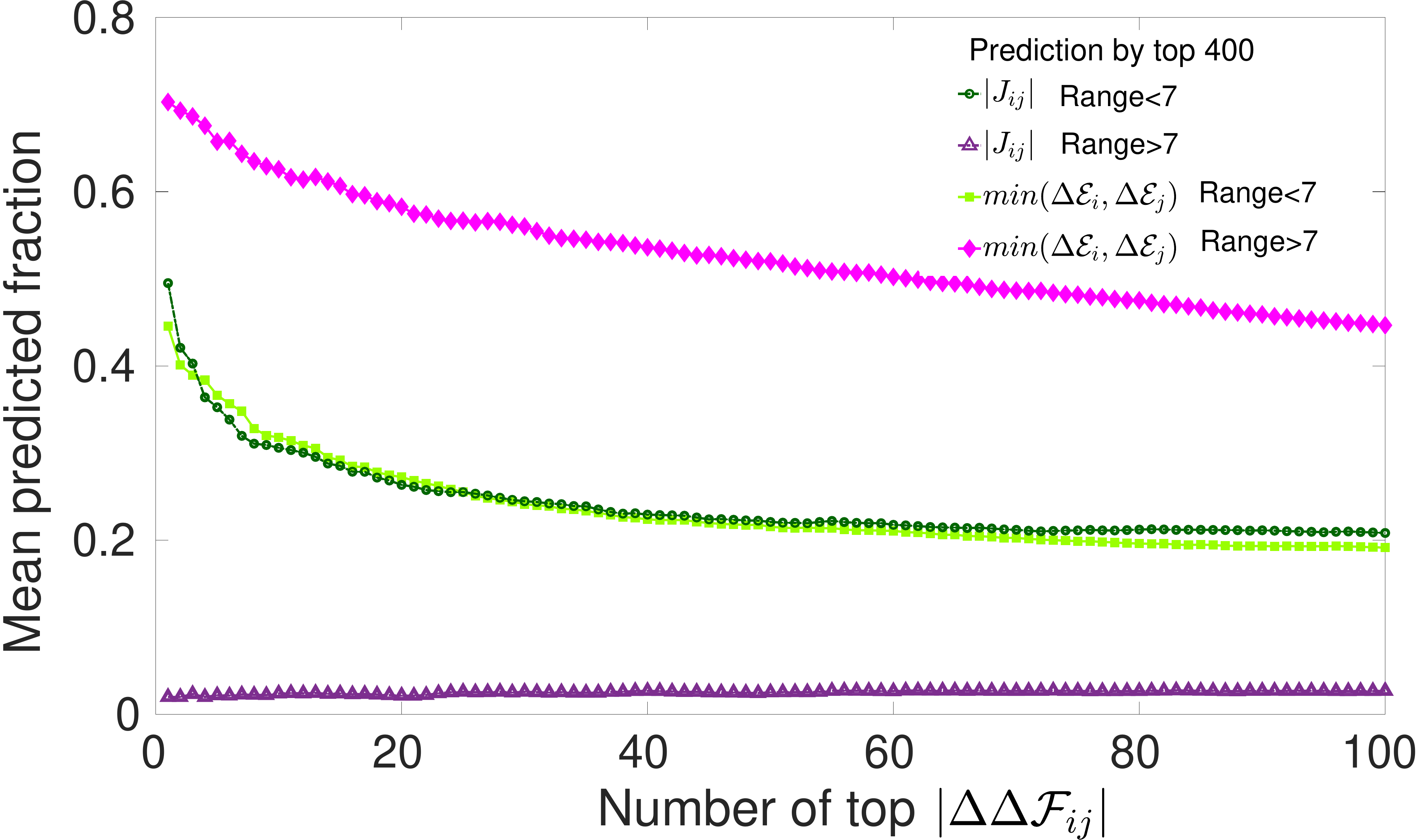}
 \includegraphics[width=0.48\textwidth]{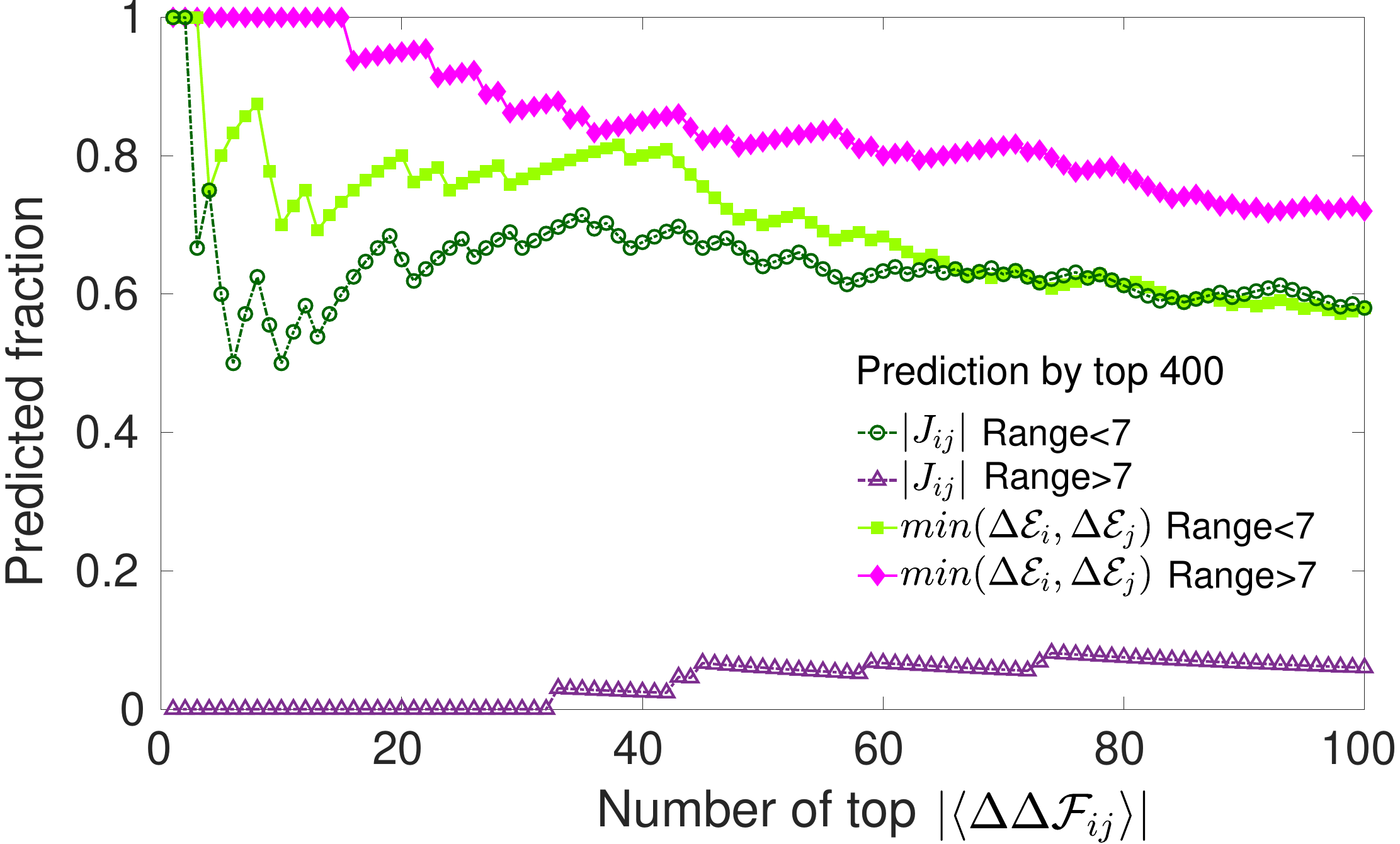}
 \caption{\textbf{Prediction of epistasis by the DCA-inferred model}. A: Same plot as in Fig.\ \ref{fig:ddf_vs_J}B (main text) where we show the fraction of top rank epistasis $|\Delta\Delta {\mathcal F}_{ij}|$ predicted by top 1000 $|J_{ij}|$, averaged over 100 configurations. In comparison to Fig.\ \ref{fig:ddf_vs_J}B, here we consider a higher number of the couplings with largest magnitude to predict epistasis: the mean predicted fraction increases both for short range and long range epistasis, yet a clear difference between their values remains. B: Same plot as Fig.\ \ref{fig:ddf_vs_J}B (main text) where we added curves for the prediction by $\min(\Delta {\mathcal E_i}, \Delta {\mathcal E_j})$ - the minimum between average single mutation costs at $i$ and $j$ - corresponding to scaling \ref{eq:scalingddf} in the main text, which describes well the trend of epistasis (see Panel A Fig. \ref{fig:epistasis}). As in Fig.\ \ref{fig:ddf_vs_J}B (main text), we rank separately long-range (> 7) and short-range (< 7) pairs of links $i$ and $j$ in terms of $|\Delta\Delta{\mathcal F}_{ij}|$ and we plot the fraction of these pairs - averaged over $100$ configurations randomly chosen - falling either into the top 400 $|J_{ij}|$ (empty symbols) or into the top 400 values of $\min(\Delta {\mathcal E_i}, \Delta {\mathcal E_j})$ (filled symbols). This second measure improves only slightly the estimation of strong short-range epistasis but it does so dramatically for long-range one. C: Same plot as B where we show the fraction of the average epistasis $\langle \Delta\Delta{\mathcal F}_{ij} \rangle$ (estimated from $1.5 \times 10^3$ randomly chosen configurations of the MSA) that one would predict either via $|J_{ij}|$ or $\min(\Delta {\mathcal E_i}, \Delta {\mathcal E_j})$. The prediction at short distance is rather accurate, with the predicted fraction reaching 1 for the maximally epistatic pairs; at long distance, signal on long-range epistasis captured by $|J_{ij}|$ is almost absent while the prediction by $\min(\Delta {\mathcal E_i}, \Delta {\mathcal E_j})$ stands out
for its precision.}
 \label{fig:ppv_min_coop}
\end{figure}

The scaling of epistasis (Eq.\ \ref{eq:scalingddf} in the main text) suggests a measure simply based on the inferred single mutation costs, i.e.\ $|\Delta\Delta {\mathcal F}_{ij}| \propto \min(\Delta {\mathcal E_i}, \Delta {\mathcal E_j})$ with $\Delta {\mathcal E}$ inferred by DCA. We have verified that this procedure improves extremely the prediction of long-range epistasis in our model for allostery in comparison to the direct evolutionary couplings $J_{ij}$, both for single configurations and for the average epistatic pattern, as shown in respectively in Panels B and C Fig.\ \ref{fig:ppv_min_coop}. 

\subsection{Simple model illustrating the failure of DCA}
\label{sec:toy_SI}
To explain the discrepancy between short-range and long-range DCA-predictions of epistasis, we resort to the simple model of Fig.\ \ref{fig:logic_gate} (main text). We assign to all the 49 functional configurations the same fitness ${\cal F}$, all the other $2^8 - 49$ configurations would not belong to the sample of optimal configurations and are taken with zero fitness, thus $\Delta {\cal F}=0$ if a mutation (single or double) results in a configuration still belonging to the optimal sample and $\Delta {\cal F} = {\cal F}$ otherwise. If we model each unit as a spin $\sigma = 0,1$, this fitness function can be mathematically written as
\be
{\cal F} = {\cal F} (\sigma_1\sigma_2+\sigma_3\sigma_4-\sigma_1\sigma_2\sigma_3\sigma_4)\cdot (\sigma_5\sigma_6+\sigma_7\sigma_8-\sigma_5\sigma_6\sigma_7\sigma_8)
\ee
i.e.\ it introduces high order couplings both at short (within groups and subparts) and long range (across subparts).\\ 
We can estimate average mutation costs by counting how frequently mutations would lead to a configuration outside of the optimal sample, yielding
\begin{equation}
\Delta\Delta {\cal F}_{12} = \Delta {\cal F}_{12} - 
\Delta {\cal F}_{1} - \Delta {\cal F}_{2} = 21/49 {\cal F} -21/49 {\cal F} - 21/49 {\cal F} = - 21/49 {\cal F}
\end{equation}
\begin{equation}
\Delta\Delta {\cal F}_{15} = 33/49 {\cal F} -21/49 {\cal F} - 21/49 {\cal F} = -9/49 {\cal F}
\end{equation}
\begin{equation}
\label{eq:toy1}
\frac{|\Delta\Delta {\cal F}_{12}|}{|\Delta\Delta {\cal F}_{15}|} = 21/9 \approx 2.3
\end{equation}
Next, by a simple likelihood maximization we infer the set of $J_{ij}$ and $h_i$ compatible with $\langle \sigma_i\rangle$ and $\langle \sigma_i \sigma_j\rangle$, single-site and pairwise frequencies of the optimal sample. We estimate $J_{12} = 1.18$ and $J_{15} = 0.40$, thus the prediction by DCA
\begin{equation}
\label{eq:toy2}
\frac{|\Delta\Delta {\cal E}_{12}|}{|\Delta\Delta {\cal E}_{15}|} = \frac{|J_{12}(2 \langle \sigma_1\rangle + 2 \langle \sigma_2\rangle - 4 \langle \sigma_1 \sigma_2 \rangle -1)|}{|J_{15}(2 \langle \sigma_1\rangle + 2 \langle \sigma_5\rangle - 4 \langle \sigma_1 \sigma_5\rangle -1)|}  = \frac{|J_{12}(-21/49)|}{|J_{15}(-9/49)|} \approx 6.9
\end{equation}
i.e.\ the DCA prediction is significantly biased towards short-range epistasis. Due to symmetry of our model, epistasis and the DCA-prediction for any combination of units in the two subparts is the same as for units $1$ and $5$; similarly, 
the result for 2 units within the same group is given by the values for units $1$ and $2$. For the remaining combinations of units, i.e.\ the ones belonging the same subpart but to different groups (e.g.\ $i=1$ and $j=3$) we obtain that epistasis is weaker compared to units within the same group
\be
\label{eq:toy3}
\frac{|\Delta\Delta {\cal F}_{12}|}{|\Delta\Delta {\cal F}_{13}|} =\frac{|-21/49{\cal F}|}{|-7/49{\cal F}|} = 3
\ee
Since each subpart can be of different type (OR gate), units from different groups (i.e.\ types) are less tightly constrained by function. The DCA-prediction does not underestimate epistasis as for units of different subparts (i.e.\ at long distance) with
\be
\label{eq:toy4}
\frac{|\Delta\Delta {\cal E}_{12}|}{|\Delta\Delta {\cal E}_{13}|} =\frac{|J_{12}(-21/49)|}{|J_{13}(7/49)|} \approx 3.5
\ee
where $J_{13} = -1.01$. From Eq.\ \ref{eq:toy1}, Eq.\ \ref{eq:toy2}, Eq.\ \ref{eq:toy3} and Eq.\ \ref{eq:toy4} it is straightforward to calculate $|\Delta\Delta {\cal E}_{13}|/|\Delta\Delta {\cal E}_{12}|\times |\Delta\Delta {\cal F}_{12}|/|\Delta\Delta {\cal F}_{13}| \approx 0.86$ and $|\Delta\Delta {\cal E}_{15}|/|\Delta\Delta {\cal E}_{12}|\times |\Delta\Delta {\cal F}_{12}|/|\Delta\Delta {\cal F}_{15}| \approx 0.33$.

\subsubsection{Feedforward neural network}
\label{sec:FF_SI}
\begin{figure}
\centering
\includegraphics[width=0.4\textwidth]{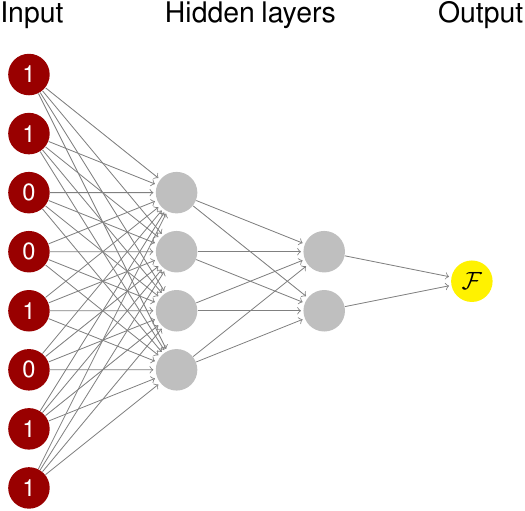}
\caption{\textbf{Graphical representation of the feedforward neural network for regression in the simple model}. The size of the input layer is 8, as the size of the system. We add two hidden layers of 4 and 2 units and the final one-unit output is 1 if the input sequence has fitness ${\cal F}$ and 0 otherwise. The activation function from one layer to the successive one is a sigmoid and the weights are dense (all units in one layer are connected to all units of the successive one).}
\label{fig:nn}
\end{figure}

To understand which machine learning tools could improve the prediction of epistasis in the simple model, we have built a feedforward neural network performing least squares regression of sequence data based on their fitness (see Fig.\ \ref{fig:nn}). For data in the training set, we provide the network with both the input sequence and the target answer, i.e.\ a label 1 (standing for fitness ${\cal F}$) or 0. We vary the size of the training set from 10\% to 80\% of the $2^8 = 256$ total sequences and we keep the remaining sequences of the sample for validation of the accuracy of prediction. We learn the weights, i.e.\ the connections between layers, which minimize the mean squared error between the output of the network and the target answers by stochastic gradient descent from a random orthogonal initialization. The 10\% of learning runs with the best performance on the training dataset reach an average training error ranging between $\sim 4\times 10^{-8}$ for a training set with 10\% of the sample (25 configurations) to $\sim 3 \times 10^{-10}$ with 80\%; the average validation error for the same runs is between $\sim 3 \times 10^{-1}$ and $\sim 2 \times 10^{-2}$ respectively. We repeated the learning with an architecture where the width of the first hidden layer is bigger than the length of input data, for instance 16 and 32. For a width of 16 hidden units, the top 10\% of trainings maintains an average accuracy on the training set of order $10^{-8}$ for the smaller training set (10\% of the sample) and of order $10^{-10}$ for the largest one (80\% of the sample); the corresponding average validation errors are $\sim 3 \times 10^{-1}$ and $\sim 4 \times 10^{-2}$. When increasing further the first layer to a width of 32, we also added a dropout (here equal to 0.3) to balance the increase of parameters to learn with the same amount of data and avoid overfitting. In this way we obtained that the training error, averaged over the 10\% best runs, was higher (from $\sim 8\times 10^{-5}$ for a training set with 10\% of the sample to $\sim 6 \times 10^{-6}$ with 80\%) but the performance on the validation set was better (respective average errors of $\sim 2\times 10^{-1}$ and $10^{-4}$). Provided that the training set is not too small, these numerical tests confirm that a trained neural network, when presented with an optimal sequence mutated at some position, can predict the value of its fitness with good accuracy in such a way as to predict $\Delta {\cal F} \sim 0$ when it still belongs to the optimal sample or $\Delta {\cal F} \sim 1$ if it does not. This ensures that also epistasis would be accurately predicted at any range.

\subsection{Inferring epistasis by Statistical Coupling Analysis}
\label{sec:epi_sca}
\begin{figure}
\includegraphics[width=0.99\textwidth]{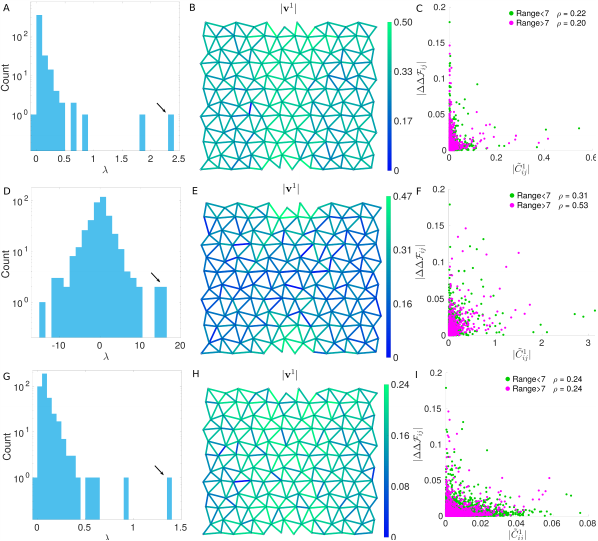}
\caption{\textbf{Measure of epistasis by SCA with conservation weight (top row A, B, C), by ICOD (central row D, E, F) and by SCA without conservation weight (bottom row G, H, I)}. 
A, D, G: Spectrum of eigenvalues $\bm{\lambda}$ of the conservation-weighted covariance (A), of the inverse off-diagonal covariance (D) and of the covariance itself (G), where the highest value $\lambda^1$ (corresponding to the first principal component $\bm{v}^1$) is highlighted by an arrow. B, E, H: Absolute values of the first principal component $\bm{v}^1$ visualized on the network (the first principal component of the conservation-weighted covariance in B, of the inverse off-diagonal covariance in E and of the covariance itself in H). C, F, I: Scatter plot of $\tilde{C}^1_{ij}$ vs epistasis with a different color code for long and short distance pairs, where $\rho$ is the Pearson correlation coefficient. $\tilde{C}^1_{ij}$ is constructed from the first top eigenvalue $\lambda^1$ and its corresponding principal component $\bm{v}^1$ of the conservation-weighted covariance in C, of the inverse off-diagonal covariance in F and of the covariance itself in I.}
\label{fig:epi_C_con}
\end{figure}

\begin{figure}
\text{{\myfont A}}\hspace{7.5cm}\text{{\myfont B}}\\
\includegraphics[width=0.45\textwidth]{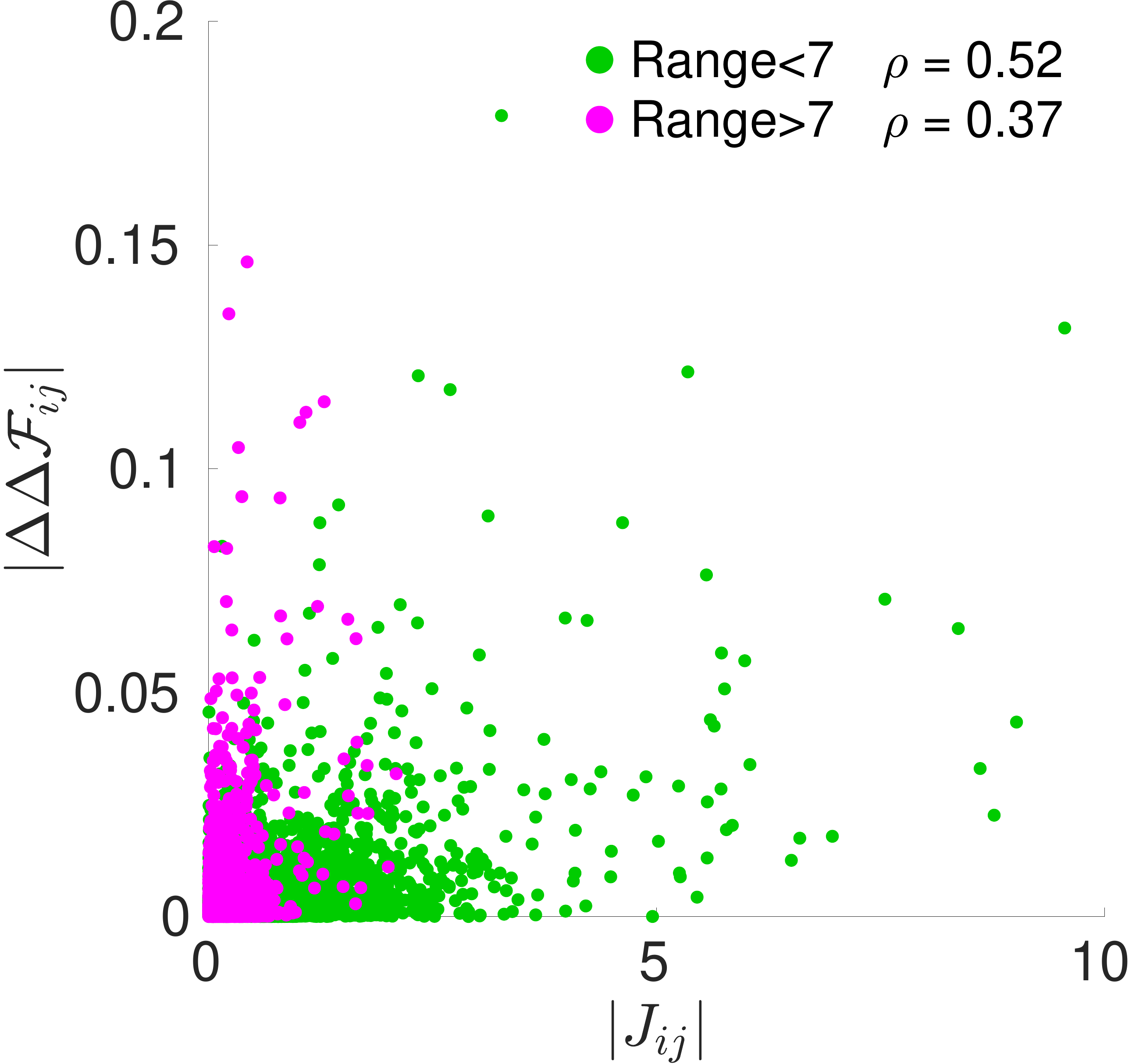}
\hspace{0.5cm}
\includegraphics[width=0.45\textwidth]{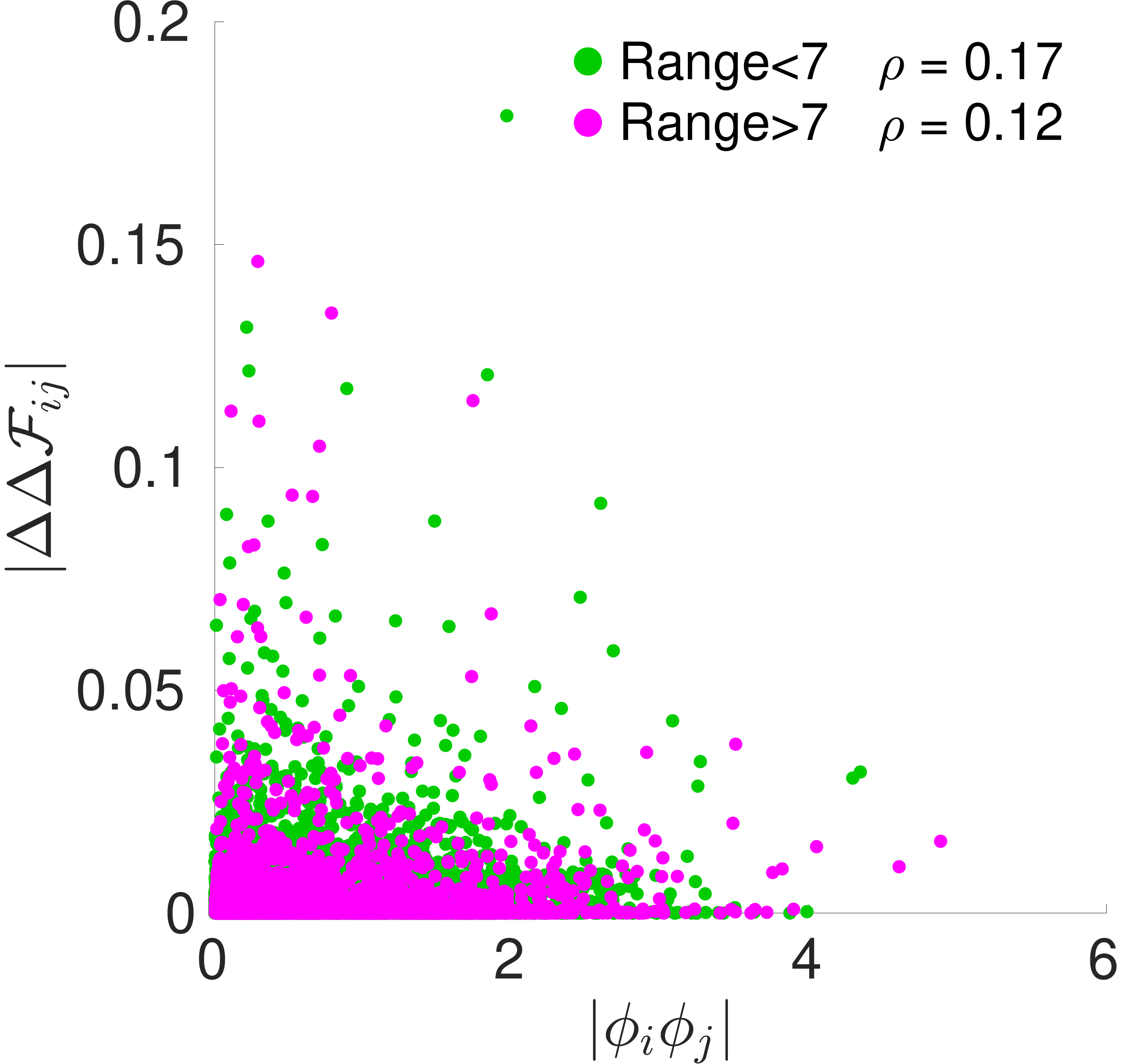}
\caption{\textbf{Epistasis measure by couplings and conservation}. A: Scatter plot of $J_{ij}$ vs epistasis, where $\rho$, the Pearson correlation coefficient, indicates a better prediction at short range. B: Scatter plot of the first mode of the conservation-only matrix $\phi_i \phi_j$ vs epistasis.}
\label{fig:epi_J}
\end{figure}

Statistical Coupling Analysis (SCA) is a principal component analysis on the covariance matrix of MSA weighted by position conservation that allows one to select the so-called ``Sectors'' \cite{Lockless99, Halabi09}. Basically, sectors consist of delocalized modes that chain together strongly co-evolving amino acids usually contiguous in the tertiary structure; they can be interpreted as basic evolutionary units that could underlie several functions including allostery \cite{Ranghanatan2003, Ranghanatan2011}. It has been recently shown \cite{Salinas18} that SCA, and in particular the first principal component only, can correctly capture a larger portion of epistasis than DCA in a deep mutational scanning experiment on the PDZ domain. Spurred by this result, we take the covariance matrix between links in our MSA of artificially evolved networks
\be
C_{ij} = \frac{1}{M} \sum_{m=1}^M \sigma_{i}^m \sigma_{j}^m - \left(\frac{1}{M} \sum_{m=1}^M \sigma_{i}^m \right)\left(\frac{1}{M} \sum_{m=1}^M \sigma_{j}^m\right)
\ee 
We apply the conservation weight prescribed by sector analysis \cite{Halabi09}, i.e.\ we consider the principal components of the matrix $\phi_i\phi_j C_{ij}$ where $\phi_i$ is a correction weighting conservation at each site. $\phi_i$ is defined as $\phi_i = \partial D_i / \partial \langle \sigma_i \rangle$ where 
\be
D_i = \langle \sigma_i \rangle\, \ln \frac{\langle \sigma_i \rangle}{\bar{\sigma}} + \big(1-\langle \sigma_i \rangle\big)\, \ln \frac{1-\langle \sigma_i \rangle}{1-\bar{\sigma}}
\ee
is the divergence of the observed occupancy of link $i$ from the background 
occupancy $\bar{\sigma}$ (see Methods) and is a measure of conservation at site $i$. 
Panel A Fig.\ \ref{fig:epi_C_con} shows the spectrum of eigenvalues $\bm{\lambda}$ of $\phi_i\phi_j C_{ij}$: the top eigenvalue $\lambda^1$ is clearly separated from the bulk (which would be shared with the spectrum of a random sample) thus incorporates information on the functional features. As in \cite{Salinas18}, we reconstruct the covariance from the top eigenmode only, $\tilde{C}_{ij}^1 = \lambda^1 v_i^1 v_j^1$, where $\bm{v}^1$ is the eigenvector corresponding to $\lambda^1$ and its structure is visualized on the network in Panel B Fig.\ \ref{fig:epi_C_con}. In Panel C Fig.\ \ref{fig:epi_C_con} we plot the absolute value of $\tilde{C}_{ij}^1$ against epistasis magnitude: it does not improve the prediction of epistasis compared to the inferred $ \Delta \Delta {\cal E}_{ij}$ or $J_{ij}$ (see Fig.\ \ref{fig:ddg_dde}, Panel A in Fig.\ \ref{fig:epi_J}) neither at short range nor at long range but, by capturing a collective mode, measures both to a more similar extent. By including the conservation weight $\phi_i$ the result is slightly improved w.r.t.\ the principal components of the uncorrected covariance, see Panels G, H, I in Fig.\ \ref{fig:epi_C_con}; on the other hand, conservation only gives a particularly poor estimation regardless of the range (Panel B Fig.\ \ref{fig:epi_J}). We have also tested the recent proposal by Wang et al.\ \cite{Bitbol19} of identifying groups of co-evolving amino acids by the top component of the inverse covariance matrix (where diagonal elements are removed before diagonalization); this method is called ICOD (Inverse Covariance Off-Diagonal). This top eigenvalue (see Panel D Fig.\ \ref{fig:epi_C_con}) corresponds to a non-local mode mainly generated by links close to the active and allosteric site (see Panel E Fig.\ \ref{fig:epi_C_con}) and correlates to long-range epistasis to a larger extent than previous methods, see Panel F Fig.\ \ref{fig:epi_C_con}.

